\documentclass[3p]{elsarticle}

\makeatletter
\def\ps@pprintTitle{%
 \let\@oddhead\@empty
 \let\@evenhead\@empty
 \def\@oddfoot{\centerline{\thepage}}%
 \let\@evenfoot\@oddfoot}
\makeatother

\usepackage{amsmath, amssymb}

\usepackage{color,soul}

\graphicspath{ {Figures/} }

\newcommand{\load}{(\tau_0-\tau_r)/(\tau_p-\tau_r)}
\newcommand{\slipvel}{2\dot u_x}
\newcommand{\peakslipvel}{2\dot u_x^p}
\newcommand{\sheartrac}{\tau}
\newcommand{\strainij}{\Delta\varepsilon_{ij}}
\newcommand{\strainxx}{\Delta\varepsilon_{xx}}
\newcommand{\strainxy}{\Delta\varepsilon_{xy}}
\newcommand{\strainyy}{\Delta\varepsilon_{yy}}
\newcommand{\strainampxx}{\delta\varepsilon_{xx}}
\newcommand{\strainampyy}{\delta\varepsilon_{yy}}

\newcommand{\xtip}{x_\mathrm{tip}}
\newcommand{\xc}{x_\mathrm{c}}

\newcommand{\Cf}{C_\mathrm{f}}
\newcommand{\CR}{C_\mathrm{R}}
\newcommand{\CS}{C_\mathrm{S}}
\newcommand{\CL}{C_\mathrm{L}}

\begin{document}

\begin{frontmatter}

\title{Dynamic fields at the tip of sub-Rayleigh and supershear frictional rupture fronts}

\author[mymainaddress,mymainaddress4]{Ilya Svetlizky  \corref{EquallyContributed}}
\cortext[EquallyContributed]{Equally Contributed}

\author[mymainaddress2,mymainaddress3]{Gabriele Albertini \corref{EquallyContributed}}
\author[mymainaddress]{Gil Cohen}

\author[mymainaddress2]{David S. Kammer \corref{mycorrespondingauthor}}
\cortext[mycorrespondingauthor]{Corresponding authors}
\ead{dkammer@ethz.ch}

\author[mymainaddress]{Jay Fineberg \corref{mycorrespondingauthor}}
\ead{jay@mail.huji.ac.il}

\address[mymainaddress]{The Racah Institute of Physics, The Hebrew University of Jerusalem, Jerusalem, Israel, 91904}
\address[mymainaddress2]{Institute for Building Materials, ETH Zurich, Zurich, Switzerland}
\address[mymainaddress3]{School of Civil and Environmental Engineering, Cornell University, Ithaca, NY 14853, USA}
\address[mymainaddress4]{School of Engineering and Applied Sciences, Harvard University, Cambridge, Massachusetts 02138, USA}

\begin{abstract}
The onset of frictional motion at the interface between two distinct bodies in contact is characterized by the propagation of dynamic rupture fronts. We combine friction experiments and numerical simulations to study the properties of these frictional rupture fronts. We extend previous analysis of slow and sub-Rayleigh rupture fronts and show that strain fields and the evolution of real contact area in the tip vicinity of supershear ruptures are well described by analytical fracture-mechanics solutions. Fracture-mechanics theory further allows us to determine long sought-after interface properties, such as local fracture energy and frictional peak strength. Both properties are observed to be roughly independent of rupture speed and mode of propagation. However, our study also reveals discrepancies between measurements and analytical solutions that appear as the rupture speed approaches the longitudinal wave speed. Further comparison with dynamic simulations illustrates that, in the supershear propagation regime, transient and geometrical (finite sample thickness) effects cause  smaller near-tip strain amplitudes than expected from the fracture-mechanics theory. By showing good quantitative agreement between experiments, simulations and theory over the entire range of possible rupture speeds, we demonstrate that frictional rupture fronts are classic dynamic cracks despite residual friction.
\end{abstract}


\end{frontmatter}

\section{Introduction}

The onset of frictional sliding occurs through a progressive failure of microcontacts forming the interface between two solids. This failure process nucleates in a localized region and propagates along the interface as a dynamic rupture front. Macroscopic sliding occurs \cite{rubinstein2004} only once these ruptures traverse the entire frictional interface. A wide range of rupture front velocities have been observed. Ruptures may propagate at a small fraction of the Rayleigh wave speed $\CR$ (slow fronts) \cite{rubinstein2004, ben-david2010science, nielsen2010} and also  asymptotically approach $\CR$. Early theoretical \cite{burridge1973,burridge1979} and numerical \cite{andrews1976} work and more recent experiments \cite{xia:2004,ben-david2010science,passelgue2013,xu2017b, mello2016,rubino2017,kammer2018,marty2019}, have shown that rupture fronts may also surpass the shear wave speed $\CS$ and reach the longitudinal speed of sound $\CL$ - the so called supershear propagation regime.

For a long time, frictional rupture fronts have been conceptually considered to be related to dynamic shear cracks, and linear elastic fracture mechanics (LEFM) has been used to study theoretically fundamental concepts of earthquake faulting \cite{book:scholz:2002,poliakov2002,bhat2007,xu2015,xu2016,xu2017}. At the sub-Rayleigh propagation regime (\textit{i.e.}, below the classical limiting velocity of cracks $\CR$ \cite{craggs:1960,kostrov:1964,freund:1972a}), one of the prominent LEFM predictions is the universal square-root singular stress fields at the crack tip. Direct experimental evidence that quantitative fracture mechanics predictions really describe frictional failure \cite{svetlizky2019} was provided by measuring explicitly these stress singularities in thermoplastic \cite{svetlizky2014,mello2016} and  rock samples \cite{kammer:2019,xu:2019}. Furthermore, LEFM has been successfully used to describe rupture-tip radiation \cite{svetlizky2016}, rupture propagation speed \cite{svetlizky2017a}, and rupture arrest \cite{kammer2015,bayart2016a,ke:2018}.

The wave radiation structure of supershear and sub-Rayleigh rupture fronts is, however, fundamentally different. The stress fields at the tip vicinity of supershear ruptures are predicted to have weaker singularities \cite{broberg1999} and, as they surpass $\CS$, shock waves have been observed to emerge \cite{xia:2004,mello2016}. In striking contrast to sub-Rayleigh rupture fronts, any attempt to describe propagation in the supershear regime requires explicit knowledge of how the singular stress fields are regularized \cite{broberg1999}. Two open questions remain in this context. First, to what extent fracture mechanics can be used to describe the near-tip fields and propagation of supershear ruptures? Second, how does regularization of the singular fields take place? While the first was partially addressed experimentally in Refs.~\cite{passelgue2013,mello2016,kammer2018}, direct measurements of frictional constitutive laws at the rupture tip are remaining a significant challenge \cite{rubino2017}. 

Here, we provide extensive experimental study of the near-tip stress fields and contact area across a wide range of rupture speeds. These include slow, sub-Rayleigh and supershear propagation regimes. We extend previous analysis of slow and sub-Rayleigh rupture fronts \cite{svetlizky2014,svetlizky2019} and show that analytical steady-state fracture mechanics solutions for thin plates describe generally well our measurements also at the supershear regime. The agreement between the measurements and the analytical fracture mechanics predictions at all speeds provides an insight into the form of the regularization and the main frictional constitutive parameters - the fracture energy $\Gamma$ and frictional strength $\tau_p$. Surprisingly, we find that both $\Gamma$ and $\tau_p$ are largely independent of rate and mode of propagation. However, we also point out apparent discrepancies between the analytical predictions and our measurements for supershear rupture fronts. We employ realistic finite-element calculations which simulate non-steady rupture propagation and take into account the third dimension (thickness) of the bodies. These simulations describe well our measurements and therefore highlight the importance of non-steady-state propagation and geometrical  effects (finite thickness) in the supershear regime. 

This paper is organized as follows. Experimental observations and the analytical model are presented and compared in section~\ref{sec:experimentsmodeling}. For completeness of the manuscript, we first review previously published analysis of slow and sub-Rayleigh rupture fronts in section~ \ref{sec:Elastic fields in the tip vicinity of sub-Rayleigh rupture fronts}. We then present measured elastic fields in the vicinity of supershear rupture fronts in section~\ref{sec:Elastic fields in the tip vicinity of supershear rupture fronts} and provide systematic comparison between the measurments and LEFM predictions for all propagation regimes in section~\ref{sec:Systematic comparison between measurements and LEFM}. The numerical model and comparison with theory and experiments are discussed in section~\ref{sec:simulations}. First we consider the effect of non-steady propagation between infinitely thin plates in section~\ref{sec:2dsim}, and then consider the effects of finite plate thickness in section~\ref{sec:3dsim}. Finally, we discuss our results and observations in section~\ref{sec:discussionconclusion}.

\section{Experiments and modeling}
\label{sec:experimentsmodeling}

\subsection{Experimental system}

We conduct an experimental study of rupture fronts propagating along a frictional interface separating two poly (methylmethacrylate) (PMMA) plates ($\rho\approx1,170~$kg/m$^3$) having the same thickness $w = 5~\mathrm{mm}$ (Fig.~\ref{fig:Fig1}). Material shear, $\CS$, and longitudinal, $\CL$, wave speeds of PMMA were obtained by measuring the time of flight of ultrasonic pulses ($\omega\sim6$~MHz), yielding $\CS=1345 \pm10$~m/s and $\CL=2700\pm10$~m/s. Due to the small wave length of the ultrasonic pulses relative to the width of the measured samples, the measured $\CL$ values correspond to plane strain hypothesis ($u_{z}=0$). Using these measured values, $\CL$ for plane stress ($\sigma_{zz}=0$) was then calculated to be $\CL = 2,333\pm10$~m/s. These measurements provide $\CR=1237\pm10$~m/s (plane stress), where $\CR$ is the Rayleigh wave speed. The optically flat interfaces of the plates ($0.1-1~\mu\textrm{m}$ in the normal direction) were allowed to wear slightly through extensive use prior to the experiments reported here. Importantly, no observable change in the roughenss of the frictional interface was observed over the stick-slip events in this work. The contacting surfaces were cleaned by isopropyl alcohol and dried for about two hours (dry interfaces from here on). We also conduct experiments in the boundary lubrication regime \cite{bayart2016b} where contacting surfaces were coated by a thin layer of lubricant, (silicon oil with kinematic viscosity $\nu \sim 100$ mm$^2$/s). In this regime the discrete asperities still bear most of the normal load, as they are not entirely immersed in the fluid layer \cite{petrova:2019}.

The two plates are carefully aligned and pressed together by an external normal force, $F_N\approx5,500N$ ($5$~MPa of nominal pressure) (see Fig.~\ref{fig:Fig1}(a)). Afterwards, shear forces, $F_S$, are applied quasi-statically until desired values of elastic energies are imposed. Slip events are subsequently nucleated on demand by inducing a slight out of plane shear (Mode III) perturbation at $x\approx0$. The imposed quasi-static force perturbations in the $z$ direction (green arrows in Fig.~\ref{fig:Fig1}(a)) are generally 2-3 orders of magnitude below $F_S$. During an event, a high speed camera (580,000 frames/s) was used to visualize the dynamic changes in the real area of contact, $A(x,t)$. 

Contact area measurements are synchronized with measurements of the 2D strain tensor, $\varepsilon_{ij}(x,t)$. $\varepsilon_{ij}(x,t)$ are measured continuously each $1\mu$s at 16-19 locations along and $3.5-4$~mm above the frictional interface [(Fig.~\ref{fig:Fig1}(a)]. When strain gauges are embedded on low modulus materials such as plastics, their presence might locally alter the strain field in their surroundings (see \cite{Ajovalasit2013} and references within). Calibration of these effects is shortly described in Ref.~\cite{Aldam2016} and further details are provided in Appendix A. 

\subsection{Frictional rupture fronts drive the onset of frictional motion }

\begin{figure}[h]
	\includegraphics{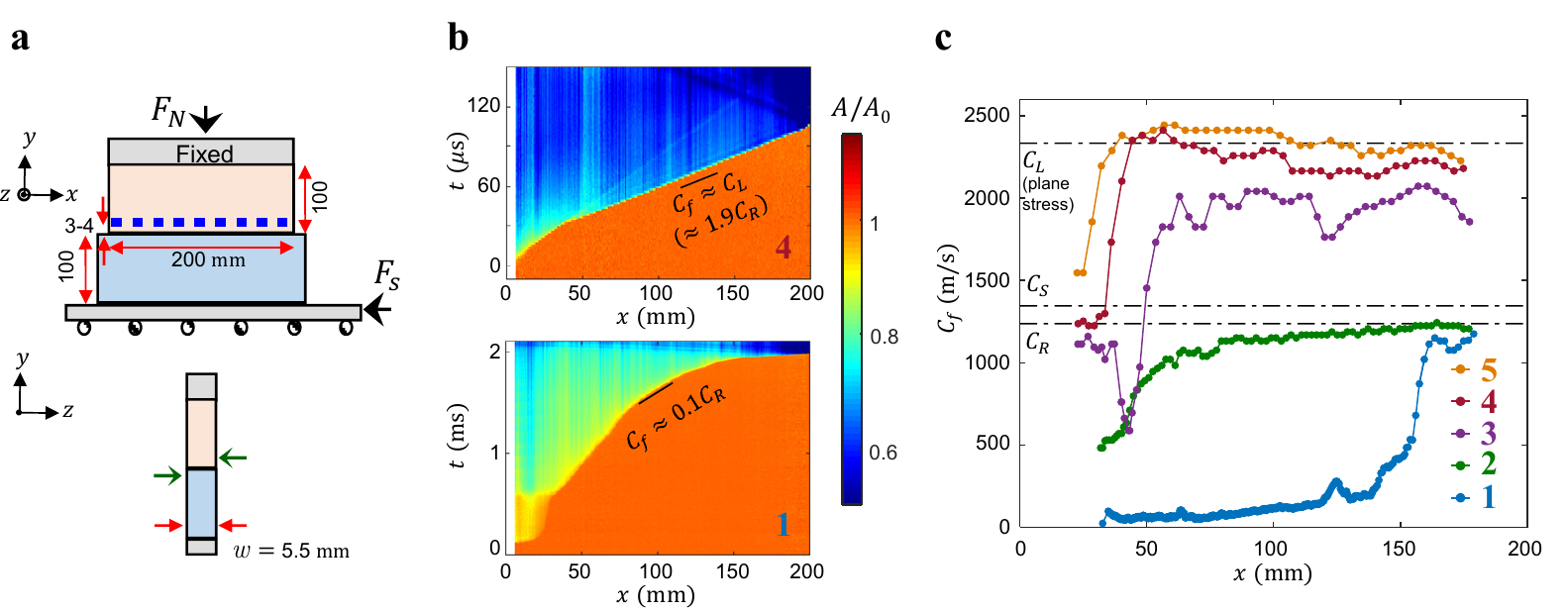}
	\centering
	\caption{Slow to supershear rupture front velocities observed experimentally. (a) Experimental system. Two poly(methylmethacrylate) (PMMA) plates are pressed together with a normal force $F_N$. The real area of contact $A(x,t)$ along the $200$~mm quasi-one-dimensional interface is measured by a method of total internal reflection at a rate of $580,000$ frames per second and averaged along the $z$ direction. In parallel, the complete two-dimensional strain tensor, $\varepsilon_{ij}$, is measured at 14-19 points (blue squares) along and slightly above the frictional interface at $1,000,000$ samples per second. Once the desired initial stresses are applied to the systems, slip events are nucleated by applying minute out of plane perturbations at $x\approx0$ (green arrows). (b) Typical examples of the short time evolution of $A(x,t)$ (normalized by $A_0(x) = A(x,t=0)$ at the time of rupture nucleation) for (bottom) a slow and (top) a supershear rupture front. (c) Typical measured rupture front velocity profiles $\Cf(x)$. Events 1 and 4 labeled in (c) correspond to the measurements of $A(x,t)$ presented in (b). Here all rupture events were obtained for the same normal load, but for different values of imposed shear stress.}
	\label{fig:Fig1}
\end{figure}

The onset of frictional motion is mediated by crack-like rupture fronts that leave in their wake significantly reduced $A$. Figure~\ref{fig:Fig1}(b) presents typical measurements of $A(x,t)$. In these examples, ruptures nucleated at $x\approx0$ and propagated in the positive $x$ direction. The boundary between regions of intact and reduced areas of contact defines the rupture tip position, $\xtip$, which is used to calculate $\Cf(x)$.  Figure~\ref{fig:Fig1}(c) demonstrates the wide range of rupture speeds observed in our experiments \cite{ben-david2010science}. These span from slow ruptures (event 1) propagating at a small fraction of the Rayleigh wave speed, $\CR$, to sub-Rayleigh ruptures that asymptotically approach $\CR$ (event 2). Rupture fronts may also surpass $\CS$ (supershear ruptures), and propagate in the range between $\sqrt{2}\CS$ (event 3) and the plane stress $\CL$ value for thin plates (event 4). Previous work employed the fracture mechanics framework and demonstrated that classical equations of motion for brittle shear cracks, derived from the energy balance at the crack tip, describes quantitatively well the velocity evolution of frictional sub-Rayleigh \cite{svetlizky2017a} and supershear rupture fronts \cite{kammer2018}. Interestingly, our experiments reveal that ruptures may also surpass $\CL$ (plane stress) by up to $5\%$ (event 5), well beyond the uncertainties in $\Cf$ and $\CL$ measurements. This will be addressed in section~\ref{sec:3dsim}. 

\subsection{Elastic fields in the tip vicinity of sub-Rayleigh rupture fronts}
\label{sec:Elastic fields in the tip vicinity of sub-Rayleigh rupture fronts}

We review now the structure of the dynamic fields at the tip of slow and sub-Rayleigh rupture fronts \cite{svetlizky2014,svetlizky2019} and address supershear ruptures in section~\ref{sec:Elastic fields in the tip vicinity of supershear rupture fronts}. In the fracture mechanics framework, rupture fronts can be modeled as propagating shear (Mode II) cracks. Crack solutions are obtained by solving two wave equations, one for longitudinal waves and one for shear waves \cite{freund1990,broberg1999} coupled by the traction free boundary conditions at the crack faces, \textit{i.e.}, $\sigma_{xy}=\sigma_{yy}=0$. 
The nominal normal stress $\sigma_{yy}^0(x)$ and the frictional resistance that opposes sliding $\tau_r(x)$ at the wake of frictional ruptures ($x-\xtip<0$) \cite{palmer1973} are taken into account by exploiting the linearity of the  governing equations. Therefore, it can be shown that the stresses at the rupture tip vicinity are dominated by 
\begin{equation}
\sigma_{ij} (r,\theta) = \frac{K_\text{II}}{\sqrt{2\pi r}}\Sigma_{ij}^\text{II}(\theta,\Cf)+
\begin{bmatrix} \sigma_{xx}^0&\tau_r\\\tau_r&\sigma_{yy}^0 \end{bmatrix}
\label{eq:subRayleigh_singular_form}
\end{equation}
where ($r,\theta$) are polar coordinates with respect to the rupture tip, $\Sigma_{ij}^{\text{II}}(\theta,\Cf)$ are known universal functions for angular dependence and the scalar $K_\mathrm{II}$ is the stress intensity factor \cite{freund1990,broberg1999}. This $1/\sqrt{r}$ singularity of the elastic fields, which essentially defines brittle fracture, is universal in the sense that its form does not depend on the geometry and outer boundary conditions. 

The energy flux through any closed contour surrounding the crack tip can be calculated. It can be shown that within the singular region, where the $K$-fields dominate, $K_\text{II}$ determines the energy flux per unit crack advance, $G_\text{II}$:
\begin{equation} G_\text{II}=\frac{K_\text{II}^2}{4(1-k^2)\mu}f_\text{II}(\Cf,k)
\label{eq:G}
\end{equation}
where $\mu$ is the shear modulus, $k=\CS/\CL$ and  $f_\text{II}(\Cf,k)$ is a known function \cite{freund1990,broberg1999} that is fairly constant for low velocities and diverges as $\Cf\rightarrow \CR$. $\CR$ is the limiting value of $\Cf$ so long as the point singularity embodied in Eq.~\ref{eq:subRayleigh_singular_form} holds \cite{svetlizky2017a}. The fracture energy $\Gamma$ is defined as the energy per unit surface area dissipated by the crack propagation. $\Gamma$ incorporates all unknown dissipation mechanisms. For a rupture to propagate, the energy flowing into the rupture tip must be balanced by the energy dissipated by creating new surface $\Gamma=G_\text{II}$.

\begin{figure}
	\includegraphics{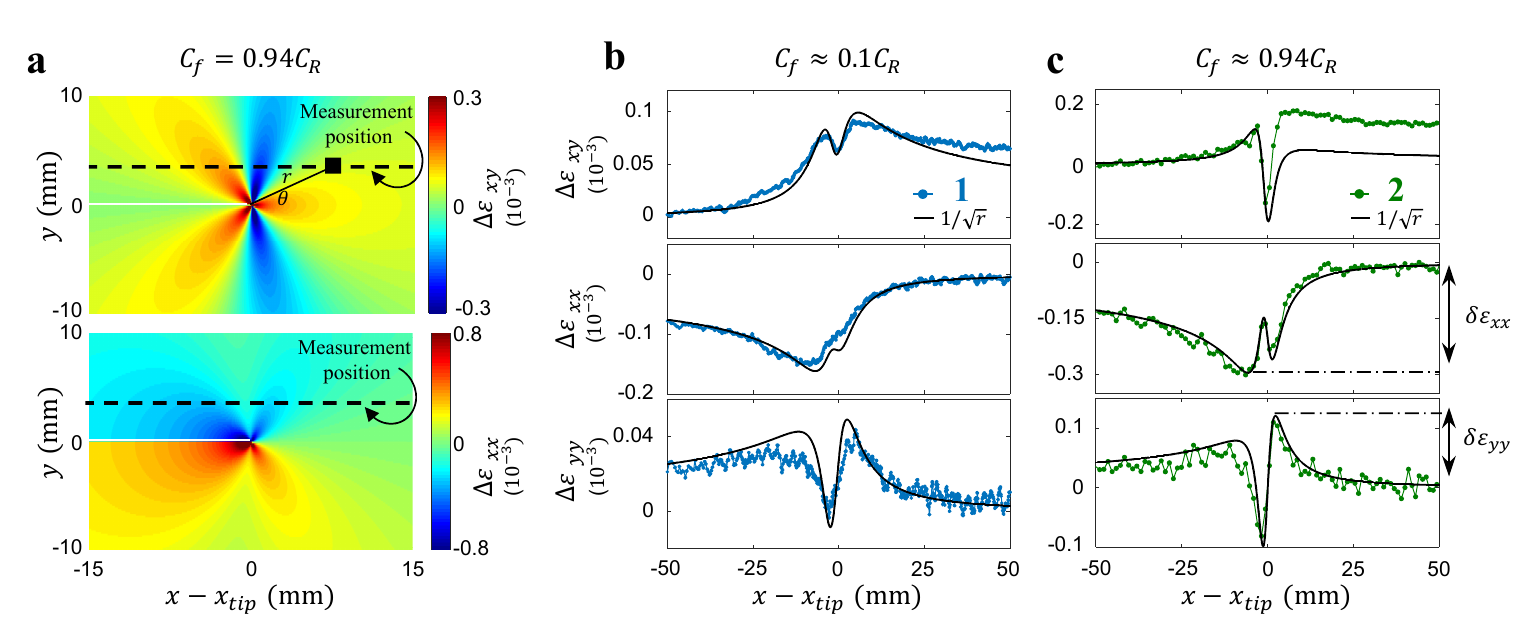}
	\centering
	\caption{The $1/\sqrt{r}$ universal form of sub-Rayleigh elastic strains. (a) Near-tip strain variations, (top) $\strainxy=\varepsilon_{xy}-\varepsilon_{xy}^r$ and (bottom) $\strainxx=\varepsilon_{xx}-\varepsilon_{xx}^0$, for $\Cf=0.94\CR$  as predicted by equation \ref{eq:subRayleigh_singular_form} and converted to strain through the linear elastic relations. Dashed lines denote the location of the strain measurements. (b) Measurements of strain tensor variations, $\strainij$, slightly above the frictional interface ($y=4$~mm), (b) for a slowly propagating rupture ($\Cf\approx0.1\CR$) and (c) rapid sub-Rayleigh rupture ($\Cf\approx0.94\CR$). The corresponding LEFM predictions are plotted in black. Note the angular dependence of the $1/\sqrt{r}$ form presented in (a) that drives the rapid oscillations is evident in the measurements. These measurements were acquired during the rupture events 1,2 presented in Fig.~\ref{fig:Fig1}(c). In both examples, $\Gamma\approx2.5$~J/m$^2$ is the sole free parameter.}
	\label{fig:SubRayleighFields}
\end{figure}

Measured strain profiles are compared with the LEFM singular predictions in Fig.~\ref{fig:SubRayleighFields}. $\varepsilon_{ij}(x,t)$ measurements at discrete locations are converted to spatial profiles, $\varepsilon_{ij}(x,t)=\varepsilon_{ij}(x-\int \Cf dt)$ by taking advantage of the high temporal resolution \cite{svetlizky2014}. For each front passage we define strain tensor variations, $\Delta\varepsilon_{ij}$, by subtracting the initial strains from $\varepsilon_{xx}$ and $\varepsilon_{yy}$ and residual strain from $\varepsilon_{xy}$. Figure~\ref{fig:SubRayleighFields}(b) shows that for slow ruptures all of the measured strain components $\varepsilon_{ij}$ agree well with the $1/\sqrt{r}$ singular form predicted by LEFM (black lines), where the only free parameter used was the value of $K_\text{II}$. Each $\varepsilon_{ij}$ measurement involved both radial and angular variations, as the strain measurements are displaced from the interface (see Fig.~\ref{fig:SubRayleighFields}(a)). $K_\text{II}$ is related uniquely to $\Gamma$ for each $\Cf$ by using Eq.~\ref{eq:G} and energy balance $G_\text{II}=\Gamma$. These measurements therefore provide the explicit value of $\Gamma$ at the interface.  

For the same constant value of $\Gamma$ nearly all of the characteristic features of $\varepsilon_{ij}$ observed at higher rupture velocities are also well described by the $1/\sqrt{r}$ form [Figure~\ref{fig:SubRayleighFields}(c)]. For example, the significant amplitude growth and the violent strain oscillations that occur when the rupture tip passes beneath the measurement point are due to the singular nature of $\Sigma_{ij}^\text{II}(\theta,\Cf)$ as $\Cf\rightarrow \CR$. The top panel in Fig.~\ref{fig:SubRayleighFields}(c) shows, however, that the spatial range in which the singular form faithfully describes  $\varepsilon_{xy}$ at high velocities is significantly smaller than that of both of the other strain components and the slow example in Fig.~\ref{fig:SubRayleighFields}(b). The far field non-singular contributions that limit the validity of Eq.~\ref{eq:subRayleigh_singular_form} are the finite prestress level $\sigma_{xy}^0$ and the stress-wave radiation that originates at the early stages of rupture acceleration \cite{svetlizky2016}. Both contributions are typically associated with faster rupture propagation.

\begin{figure}[h]
	\includegraphics{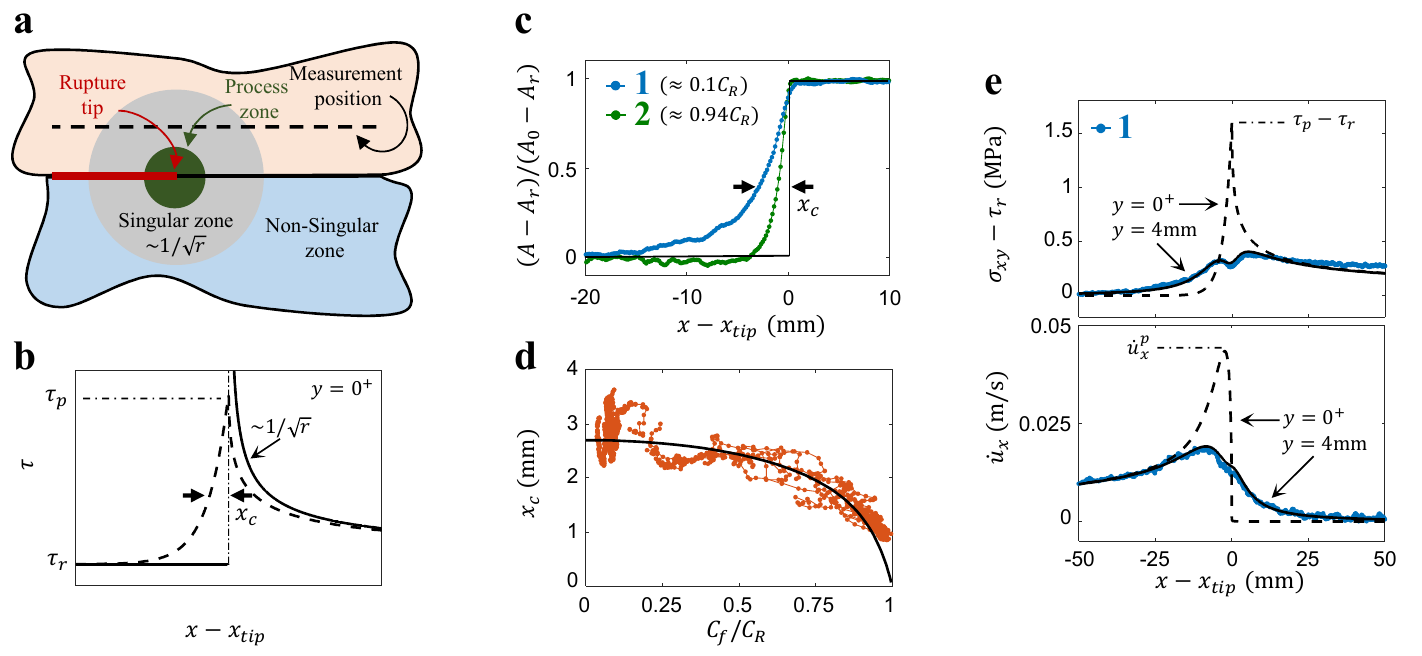}
	\centering
	\caption{ Regularization of the elastic singular fields at the rupture tip. (a) Stresses at the vicinity of the crack tip have a universal $1/\sqrt{r}$ singular form (gray region). Singular stresses are regularized in the process zone (green region), where dissipation takes place. (b) Schematic drawing of a nonsingular cohesive zone model (dashed line) in which the shear stress is reduced exponentially behind the crack tip once the peak strength, $\tau_p$, is reached. $\xc$ represents the size of the process zone, the scale at which the singular fields are regularized. Far ahead of the crack tip ($x-\xtip\gg \xc$) the solution matches the $1/\sqrt{r}$ singular form (black line). (c) The normalized real area of contact, $A$, is plotted relative to the rupture tip position for the two rupture events shown in Fig.~\ref{fig:Fig1}(c) and Fig.~\ref{fig:SubRayleighFields}. $A_0$ and $A_r$ are the initial (prior to rupture arrival) and residual (after rupture passage) values of $A$, respectively. $\xc$ is the length scale corresponding to a 60$\%$ reduction of $A$ and represents an estimate of the process zone size defined in (b). (d) $\xc$ contracts as $\Cf\rightarrow \CR$. Black line - LEFM prediction with a sole free parameter $\xc(\Cf=0)$. (e) Measurements of the shear stress variation (top) and particle velocity $\dot{u}_x=-\varepsilon_{xx}\Cf$ \cite{svetlizky2014} at the measurement points located $4$~mm above the interface. The exponential cohesive zone model at the interface is defined by two parameters, $\Gamma$ and $\xc$. $\xc$ is estimated from measurements of $A$ in (c). $\Gamma$ was measured in Fig.~\ref{fig:SubRayleighFields}. Interface (dashed lines) and off-interface (solid lines) predictions of the cohesive zone model.}
	\label{fig:Xc}
\end{figure}

The infinite stresses at the rupture tip, predicted by the singular LEFM solutions, are naturally regularized in the process (cohesive) zone in the near vicinity of the tip. In the ``small scale yielding'' \cite{Rice1968} approximation, the region where the square-root singular form dominates, therefore, is interpreted as an ``intermediate asymptotic'' region separating the ``inner'' scales of dissipation from the ``outer" region where non-singular contributions can not be neglected [see Fig.~\ref{fig:Xc}(a)]. Regularization of rupture tip singularities is not expected to be universal, as various dissipative processes may take place. Direct measurements of the constitutive evolution law of friction are generally impeded by the singular nature of the fields. An intensive and on-going effort is directed towards formulation of constitutive laws which endeavors to address the rate and history dependence of the frictional resistance. While ignoring the intricate details of friction, simple cohesive zone models, with the advantage of being analytically tractable, provide insight into the dynamics near the rupture tip  \cite{palmer1973,poliakov2002,samudrala2002,kammer:2019}. These models often assume that weakening of the local frictional resistance, $\tau$, is initiated once the shear stress has reached a finite peak strength, $\tau_p$. $\tau$ continuously decrease from $\tau_p$ to $\tau_r$ with the spatial position, according to a prescribed spatial stress profile, $\tau(x)=(\tau_p-\tau_r)\cdot\widetilde{\tau}((x-\xtip)/\xc)+\tau_r$ [see Fig.~\ref{fig:Xc}(b)], where $\xc$ is defined to be the cohesive zone size. The problem is closed by the universal boundary conditions dictated by the singular $K-$fields, \textit{i.e.}, far ahead of the crack tip the solution matches the square-root singular form, $\sigma_{xy}(x-\xtip \gg \xc, y=0)\rightarrow K_\text{II}/\sqrt{2\pi (x-\xtip)}$ \cite{samudrala2002}. Therefore,  $(\tau_p-\tau_r)$, $\xc$ and $\Gamma$ are related through Eq.~\ref{eq:G} by 
\begin{equation}
\Gamma=(\tau_p-\tau_r)^2 \xc\frac{\widetilde{G}(\Cf,k,\widetilde{\tau}(\xi))}{2\pi(1-k^2)\mu}
\label{eq:Cohezive Zone}
\end{equation}
where $\widetilde{G}(\Cf,k,\widetilde{\tau}(\xi))=f_{\text{II}}(\Cf,k)[\int_{-\infty}^{0}\frac{\widetilde{\tau}(\xi)}{\sqrt{-\xi}}d\xi]^2$. Even for the simple case where $\Gamma$ and $\tau_p$ are rate independent, rewriting Eq.~\ref{eq:Cohezive Zone} to  $\xc(\Cf)=\xc(\Cf=0)/f_\text{II}(\Cf,k)$, demonstrates the $\xc$ dependence on the rupture speed, often referred to as the ``Lorentz" contraction of $\xc$ (note that $f_\text{II}\rightarrow \infty$ as $\Cf \rightarrow \CR$).

Strain measurements in our experiments, which are slightly displaced from the frictional interface, are described by $1/\sqrt{r}$ singular solutions (Fig.~\ref{fig:SubRayleighFields}) and essentially insensitive to the illusive processes at the rupture tip. Measurements of the real contact area, however, may shed light on how these singular fields are regularized, since these measurements, by definition, take place {\em on} the interface. Figures~\ref{fig:Xc}(c) shows that $A(x,t)$ decreases gradually behind the propagating rupture tip, in clear contrast to idealized singular cracks for which an abrupt $A(x,t)$ reduction is expected. The length scale over which $A(x,t)$ is reduced provides an estimate of the cohesive zone size, $\xc$ \cite{svetlizky2014}.  $\xc$ is not constant [Figure~\ref{fig:Xc}(c,d)], and the systematic contraction of $\xc$ with increasing $\Cf$ is well described by LEFM (Eq.~\ref{eq:Cohezive Zone}) where $\xc(\Cf=0)$ is the only free parameter.  

The knowledge of $\Gamma$ and  $\xc(\Cf=0)$ and assuming $\widetilde{\tau}(\xi)=e^{\xi}$ (alternative assumptions regarding the $\widetilde{\tau}(\xi)$ functional form do not significantly affect the results) enable us to estimate the elusive but long sought after constitutive parameters that characterize the dissipative processes and material properties at the extreme conditions that take place near the rupture tip. The peak shear strength $\tau_p$ is inferred directly from Eq.~\ref{eq:Cohezive Zone}, while the maximal slip velocity $2\dot{u}_x$ and critical slip distance, $d_c$ are found by calculating explicitly the dynamic fields [Fig.~\ref{fig:Xc}(e)]. The corresponding local residual and static friction coefficients are estimated to be $\mu_r \approx 0.45$ and $\mu_s \approx 0.77$, respectively. Figure~\ref{fig:Xc}(e) reveals that measurements performed slightly above the frictional interface do not reflect any of the cohesive zone dynamics or properties. Neither the peak strength nor the cohesive zone size can be inferred directly from these off-interface measurements. This is especially true for $\Cf\rightarrow \CR$, since $\xc \rightarrow 0$ in this regime.  Measurements at finite distances from a frictional interface should, therefore, be interpreted with extreme caution as the divergence of the near-tip fields may result in erroneous conclusions. Credible measurements of the interface properties are particularly challenging as they can only be obtained if measurements are performed at distances from the rupture tip that are much smaller than $\xc$. It is progressively harder to meet this requirement as $\xc$ contracts with the rupture velocity.

\subsection{Elastic fields in the tip vicinity of supershear rupture fronts}
\label{sec:Elastic fields in the tip vicinity of supershear rupture fronts}

\begin{figure}[h]
	\includegraphics{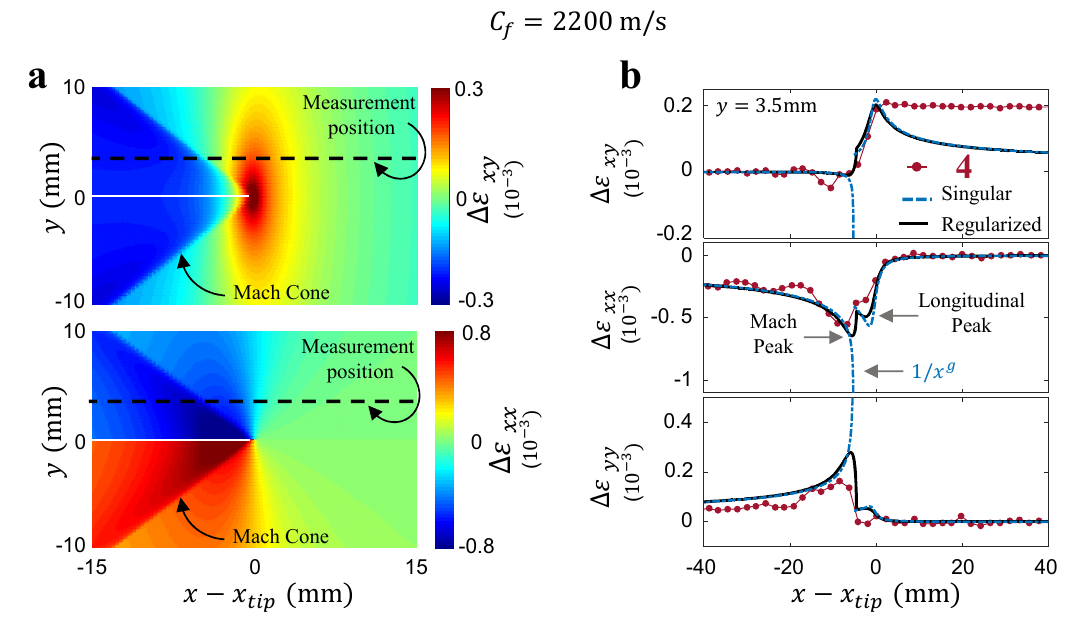}
	\centering
	\caption{ Supershear rupture fronts -- measurements along the Mach cone provide an estimate of the shear strength peak $\tau_p-\tau_r$. (a) Strain variations, (top) $\strainxy$ and (bottom) $\strainxx$, surrounding the rupture tip predicted by the cohesive zone model (see text for more details).  Dashed lines denote the location of the strain measurements. Note the fundamental difference in the structure of fields during supershear and sub-Rayleigh [Fig.~\ref{fig:SubRayleighFields}(a)] rupture propagation. (b) Measurements of strain tensor variations, $\strainij$, slightly above the frictional interface (measurement position is denoted by the dashed lines in panel a). These measurements were acquired during rupture event 4 presented in Fig.~\ref{fig:Fig1}(c). The divergence of the elastic fields along the Mach cone, predicted by the singular solution (blue dashed lines), is regularized by the cohesive zone model (black solid lines). Measurements of the Mach peak amplitudes (defined in the figure), therefore, provide a direct estimate of $\tau_p-\tau_r$. The local rupture speed is $C_f=2200$~m/s, which results in $g\approx0.42$. For the cohesive zone model we use $\Gamma\approx2.5$~J/m$^2$ and $\tau_p-\tau_r=1.62$~MPa [$(\tau_p-\tau_r)/2\mu\approx0.38\cdot10^{-3}$], as were inferred from the strain and contact area measurements at the sub-Rayleigh regime.}
	\label{fig:SuperShearFieldsSingular}
\end{figure}

\begin{figure}[h]
	\includegraphics{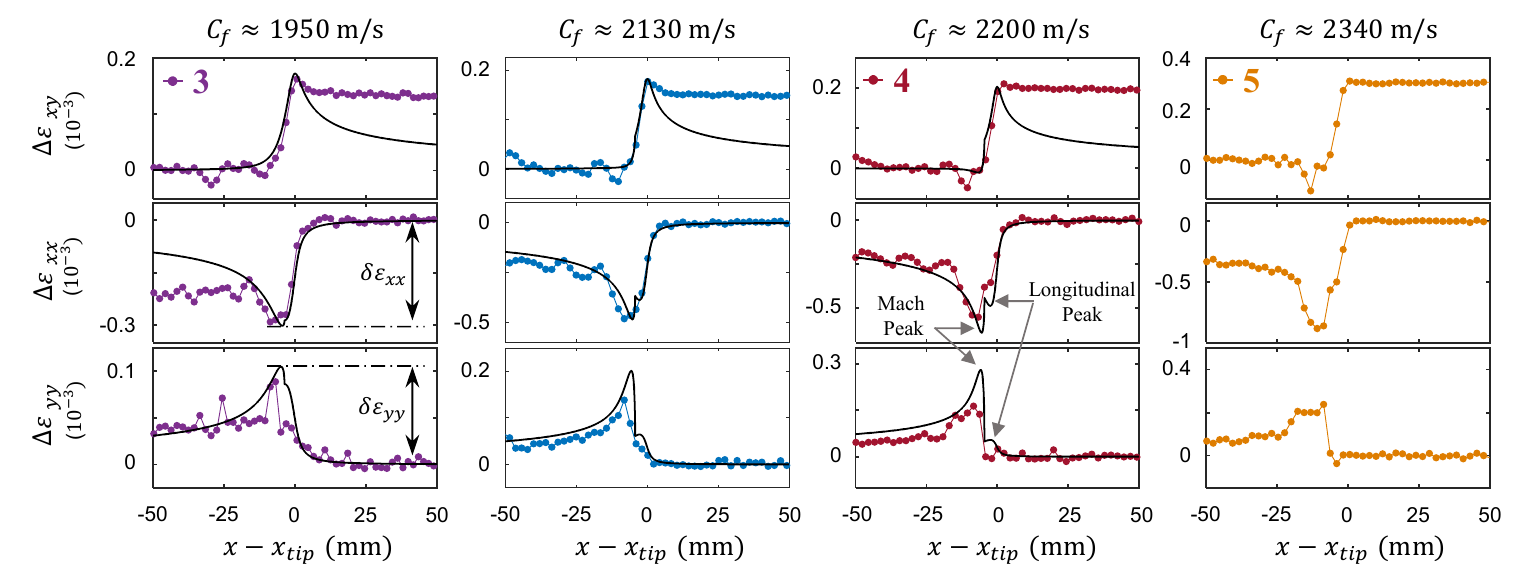}
	\centering
	\caption{Measured elastic strain fields at the supershear rupture tip vicinity. Measurements of  $\strainij$ acquired during supershear ruptures propagating at different speeds [rupture speed profiles are presented in Fig.~\ref{fig:Fig1}(c)]. Note the increasing strain amplitudes ($\strainampxx$ and $\strainampyy$) with increasing $\Cf$. The corresponding predictions of LEFM coupled to the cohesive-zone model (following Section~6.3 in Ref.~\cite{broberg1999}) are plotted in black with no adjustable parameters; $\Gamma$ and $\tau_p$ were determined from measurements performed in sub-Rayleigh propagation (Fig.~\ref{fig:SubRayleighFields} and Fig.~\ref{fig:Xc}) and specified in the caption of Fig.~\ref{fig:SuperShearFieldsSingular}. Since $\Cf$ exceeds $\CL$ (plane stress) in event 5, no comparison with the model is provided (see main text).}
	\label{fig:SuperShearFields}
\end{figure}

Early steady-state analytical solutions \cite{Freund1979,broberg1999} demonstrated substantial differences between the sub-Rayleigh and supershear cracks that propagate at speeds above the shear wave speed. For supershear cracks, the contribution of the longitudinal waves to the stresses forms a singularity at the crack tip $\sigma_{ij} \sim 1/r^g$. The singular exponent $g$ depends on the crack speed ($g(\Cf) \le 1/2$), in striking contrast to sub-Rayleigh cracks. Furthermore, as the crack tip speed exceeds $\CS$, shear waves radiated at the crack tip always trail behind and form shock waves at all speeds $\Cf \neq \sqrt{2}\CS$ [Fig.~\ref{fig:SuperShearFieldsSingular}(a)]. These shock waves contain singularities in the stress, strain and material velocities  along the Mach cone, and the singular exponent matches the singularity at the crack tip - $g$. Figure.~\ref{fig:SuperShearFieldsSingular}(b) demonstrates, however, that in contrast to the predictions of the singular solutions (blue dashed lines), the measured amplitudes of the shock waves are finite. 
Shock waves, propagating at considerable distances from the crack tip, contain therefore detailed information about the elusive fracture processes at the crack tip \cite{dunham:2005,mello2010,mello2016}, which are typically obscured in the sub-Rayleigh regime. How strain measurements in the supershear regime can be employed to infer $\tau_p-\tau_r$ values is one of the key results of this work and will be demonstrated in what follows.

In order to compare measured strain profiles with the fracture mechanics predictions we consider a direct extension of the sub-Rayleigh cohesive zone model (Sec.~\ref{sec:Elastic fields in the tip vicinity of sub-Rayleigh rupture fronts}) to the supershear regime (Section~6.3 in Ref.~\cite{broberg1999}). $(\tau_p-\tau_r)$, $\xc$ and $\Gamma$ are related through the functional form provided by Eq.~\ref{eq:Cohezive Zone}, once the supershear analogue of $\widetilde{G}(\Cf,k,\widetilde{\tau}(\xi))$ is substituted (Eq.~6.3.69 in Ref.~\cite{broberg1999}). Therefore, for any given $\Cf$, $\Delta\varepsilon_{ij}$ are calculated explicitly [Fig.~\ref{fig:SuperShearFieldsSingular}(a)], when provided the elastic constants ($\mu$ and $k$) and cohesive zone properties [$\Gamma$, $\tau_p$ and $\widetilde{\tau}(\xi)$]. Note that, by introducing a cohesive zone, the requirement for a finite and positive energy flux into the crack tip $G_\text{II}$ is fulfilled \cite{broberg1999}. This is in contrast to an unrealistic consequence of the singular description of supershear cracks that predicts vanishing $G_\text{II}$ for $\Cf\not=\sqrt{2} \CS$. 

It was shown \cite{broberg1999} that the regularization for the shock wave singularity naturally emerges once a cohesive zone is introduced and the stress divergences at the crack tip are regularized. Figure~\ref{fig:SuperShearFieldsSingular}(b) shows that the predictions of the regularized solution (black solid lines) reasonably agree with the measured $\Delta\varepsilon_{ij}$ profiles with no adjustable parameters; we exploit $\Gamma$ and $\tau_p-\tau_r$ values inferred from sub-Rayleigh rupture events and keep, for simplicity, $\widetilde{\tau}(\xi)=e^\xi$. In particular, the regularized solutions, in contrast to the singular predictions, describe rather well the finite Mach peak amplitudes. This result is further demonstrated in Fig.~\ref{fig:SuperShearFields} where additional spatial strain profiles at various supershear rupture speeds are plotted. As the rupture speed increases above $\sqrt{2}\CS$ ($\sim1900$~m/s), Mach cones are observed to form and grow in amplitude in both the $\varepsilon_{xx}$ and $\varepsilon_{yy}$ components [see definition in Fig.~\ref{fig:SuperShearFieldsSingular}(b)]. The contributions of longitudinal waves, however, differ for the two components. The ``longitudinal peak" increases (decreases) in size for the $\varepsilon_{xx}$ ($\varepsilon_{yy}$) components with increasing rupture speeds. 

The top panels in Fig.~\ref{fig:SuperShearFields} demonstrate that the solution does not capture $\Delta\varepsilon_{xy}$ ahead of the rupture tip ($x-\xtip>0$). This should be expected as in this region, similarly to the high pre-stress sub-Rayleigh examples [Fig.~\ref{fig:SubRayleighFields}(c)], the assumption of vanishingly small remote stresses ($\sigma_{xy}^0\ll\tau_p$) does not hold. We will address this point again in what follows (Sec.~\ref{sec:2dsim}) when we consider numerical simulations. 

For completeness, we present $\Delta\varepsilon_{ij}(x)$ measurements for a rupture that surpasses the estimated plane stress $\CL$ value [example 5 in Fig.~\ref{fig:Fig1}(c) and Fig.~\ref{fig:SuperShearFields}]. As these measurements do not show any dramatic difference compared to the examples at lower speeds they suggest that the plane stress $\CL$ does not necessarily provides the correct limiting velocity in our experiments. In Sec.~\ref{sec:3dsim} we will argue that the finite width of the samples plays an important role and the thin plate approximation which we have adopted in this discussion has a limited range of applicability. 

\subsection{Systematic comparison between measurements and LEFM}
\label{sec:Systematic comparison between measurements and LEFM}

\begin{figure}
	\includegraphics{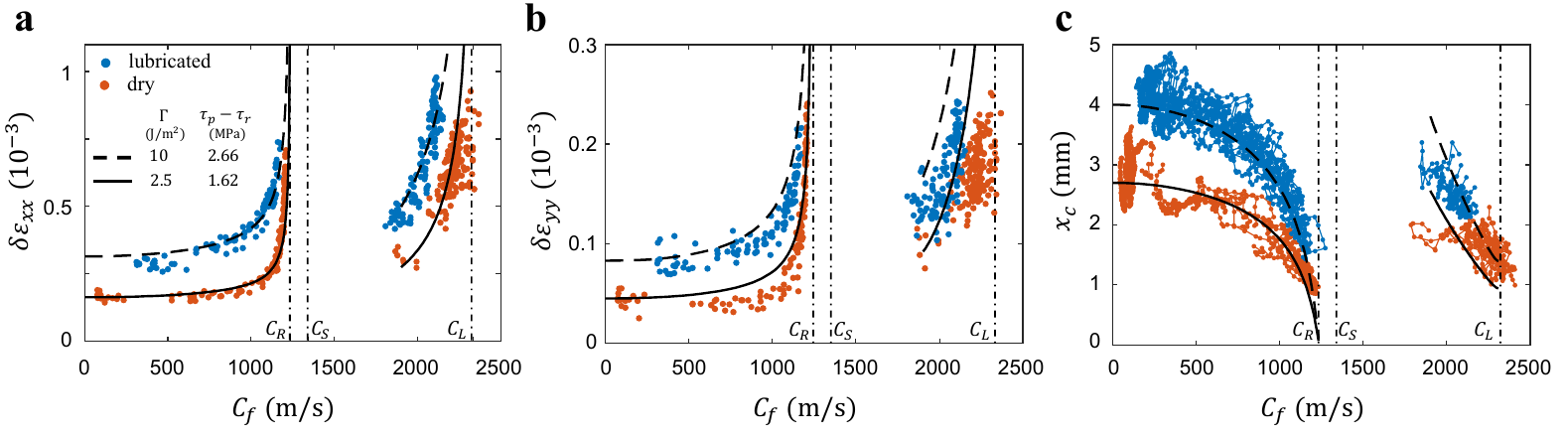}
	\centering
	\caption{Systematic comparison of measurements to LEFM predictions for rupture speeds that span the entire range of allowed rupture velocities, from slow to supershear speeds. Elastic strains, $\varepsilon_{xx}$ and $\varepsilon_{yy}$, measured at $y=3.5-4$~mm  are characterized by their amplitudes, $\strainampxx$ (a) and $\strainampyy$ (b), as defined by the double arrows in Fig.~\ref{fig:SubRayleighFields} and Fig.~\ref{fig:SuperShearFields}. Measurements of $\xc$ [see definition in Fig.~\ref{fig:Xc}(c)] are plotted in (c). Colors represent experiments with dry (orange) and boundary-lubricated interfaces (blue). LEFM predictions, for rate independent values of $\Gamma$ and $\tau_p-\tau_r$, are plotted in black. $\Gamma$ and $\tau_p-\tau_r$ values (see legend) were determined by fitting $\strainampxx$ and $\xc$ at sub-Rayleigh rupture speeds ($\Cf<\CR$) and then used to compare with measurements within the supershear regime ($\Cf>\CS$). } 
	\label{fig:AmpVsCf}
\end{figure}

We now consider a systematic comparison between the cohesive zone model and a compilation of $\sim80$ rupture events that span a wide range of speeds (Fig.~\ref{fig:AmpVsCf}). These include slow ruptures ($\Cf\ll \CR$), ruptures that asymptotically approach $\CR$ and ruptures that surpass $\CS$. No ruptures have been observed to propagate a substantial distance at $\CS<\Cf<\sqrt{2}\CS$, although rapid acceleration through this range of velocities may occur [see Fig.~\ref{fig:Fig1}(c)]. This observation is consistent with previous energy balance considerations of propagating supershear ruptures \cite{burridge1979,broberg1994,obrezanova:2003,kammer2018}. To extend the generality of our results, we supplement the dry friction experiments (orange data points in Fig.~\ref{fig:AmpVsCf}) with analysis of experiments in the boundary lubricated regime (blue data points in Fig.~\ref{fig:AmpVsCf}). 

We concentrate here on the rupture speed dependence of both, the strain amplitudes, $\delta\varepsilon_{xx}$ and $\delta\varepsilon_{yy}$ [Fig.~\ref{fig:AmpVsCf}(a,b)], and $\xc$ [Fig.~\ref{fig:AmpVsCf}(c)]. $\delta\varepsilon_{xx}$ and $\delta\varepsilon_{yy}$ are defined in Fig.~\ref{fig:SuperShearFields}, and $\xc$ is defined in Fig.~\ref{fig:Xc}(c). We limit the observation region to $100$~mm~$<x<140$~mm, where no strong heterogeneities of the frictional interface are present. While both $\delta\varepsilon_{xx}$ and $\delta\varepsilon_{yy}$ rapidly increase, $\xc$ contracts as $\CR$ is approached, jumps to a finite value beyond $\CR$ and then contracts again when $\CL$ is approached. This rather general behavior of $\xc$ was previously discussed in Refs.~\cite{kubair2002,bhat2007}. 

We use the slow rupture regime ($\Cf \ll \CR$) to determine cohesive zone properties \cite{svetlizky2014,svetlizky2017a}. First, $\Gamma$ is determined by comparing $\delta\varepsilon_{ii}$ with the singular LEFM solution (Eq.~\ref{eq:subRayleigh_singular_form}) and using Eq.~\ref{eq:G}. The value of $\tau_p-\tau_r$ is then estimated through $\xc(\Cf\approx0)$ measurements [Fig.~\ref{fig:AmpVsCf}(c)] by using Eq.~\ref{eq:Cohezive Zone}. The inferred values of $\Gamma$ and $\tau_p-\tau_r$, for both the dry and boundary lubrication interfaces, are specified in the legend in Fig.~\ref{fig:AmpVsCf}(a). Interestingly, the addition of a thin layer of lubricant significantly increases $\Gamma$. The inferred increase in $\tau_p-\tau_r$, however, is smaller in extent, and thereby suggests an increase in slip weakening distance $d_c$. We note that the increase in $\tau_p - \tau_r$ is largely due to the decrease of dynamic friction $\tau_r$ (not shown here) and not due to an increase in $\tau_p$. These observations might suggest that once slip is initiated, the lubricant trapped within a rough layer of contacts is released and facilitates slip, and thereby reducing $\tau_r$ and increasing $d_c$ (for more details see Ref. \cite{bayart2016b}).

The values of $\Gamma$ and $\tau_p-\tau_r$ that were obtained in the slow regime provide excellent descriptions of our measurements through the entire sub-Rayleigh range ($\Cf<\CR$) (solid and dashed black lines in Fig.~\ref{fig:AmpVsCf}). $\delta\varepsilon_{ii}$ predictions based on the singular and regularized solutions, for $\Cf<\CR$ do not show a significant difference, as strains are measured at heights above the frictional interface which are comparable to $\xc$. Furthermore, Fig.~\ref{fig:AmpVsCf} demonstrates that the applicability of LEFM extends to the supershear regime. For the same values of $\Gamma$ and $\tau_p-\tau_r$, the cohesive zone model generally describes our measurements well over the whole range of rupture speeds. The simplicity of this result is surprising; both $\Gamma$ and $\tau_p-\tau_r$ are largely rate independent for both the dry and lubricated interfaces. 

We note that, while for $r>\xc$ and $\Cf<\CR$ predictions based on the singular and regularized solutions show a close resemblance, this is no longer true for $\Cf>\CS$. The shock waves are sensitive to the cohesive zone properties. In particular, the shock wave amplitudes diverge with increasing $\tau_p-\tau_r$. Therefore, the agreement between the $\delta\varepsilon_{ii}$ measurements and the cohesive model at the supershear regime provides a further independent validation of our $\tau_p-\tau_r$ estimates. Furthermore, current analysis of supershear ruptures demonstrates that strain measurements performed away from the rupture tip can provide a reliable tool for estimating $\tau_p-\tau_r$ values in opaque materials, where measurements of $A$ (and therefore $\xc$) are impossible.

Despite the impressive agreement shown in Fig.~\ref{fig:AmpVsCf}, some apparent discrepancies between measurements and fracture mechanics predictions within the supershear regime can be observed. The cohesive zone model consistently overestimates the measured strain amplitudes as $\Cf\rightarrow\CL$. This will be addressed in Sec.~\ref{sec:3dsim}. We will argue that for $\Cf>\CS$ the infinitely thin plate approximation adopted to obtain analytical predictions is limited and a 3D description is necessary to account in detail for the measurements.

\section{Comparison to dynamic rupture front simulations}
\label{sec:simulations}

We use finite-element simulations to further analyze our experimental results and address the discrepancies between our measurements and the analytical LEFM predictions in the supershear regime. In particular, we address the following observations: (1) Ruptures may surpass $\CL$ (plane stress) by up to $5\%$ (Fig.~\ref{fig:Fig1}). (2) Apparent discrepancies appear in $\strainampxx$ and $\strainampyy$ (Fig.~\ref{fig:AmpVsCf}) as $\Cf\rightarrow\CL$. Simulations allow us to verify the \emph{strong} assumptions on the analytical models namely steady-state rupture propagation and two dimensional geometry. 

The structure of this section is as follows. Finite-element simulations are briefly presented in section~\ref{sec:Numerical_model}. In section~\ref{sec:2dsim} we consider rupture propagation between infinitely thin plates and address the applicability of the steady-state analytical solutions to accelerating ruptures. In section~\ref{sec:3dsim} we discuss the effects of the sample thickness and present full 3D numerical simulations together with their systematic comparison with the experiments.

\subsection{Numerical model}
\label{sec:Numerical_model}

Simulations model two solid bodies that are in contact along a flat frictional interface. The onset of sliding occurs as a localized failure that propagates spontaneously along the interface. Our numerical setup with the parameter choice described below is quantitatively equivalent to the (dry interface) experimental system with identical length scales, stress levels, material and interface properties. 

We apply a dynamic finite-element method to solve the elastic wave equations of the solids. Time integration is computed with a Newmark-$\beta$ explicit scheme, which uses a predictor-corrector formulation. We apply a lumped mass matrix for computational efficiency and a regular mesh with up to $4000$ elements along $x$-axis of the interface. The solid blocks are of length $L=0.2~\mathrm{m}$, height $H=L/2$ and various widths $w$ (see section~\ref{sec:3dsim}). A particularly large simulation used a value of $L=0.1~\mathrm{m}$ to reduce computational cost. We applied linear elastic material properties with elastic modulus $E=5.65~\mathrm{GPa}$, Poisson's ratio $\nu=0.33$ and density $\rho=1160~\mathrm{kg/m}^3$, which yield similar wave speeds as those measured in the experimental system. Viscoelastic properties of PMMA are neglected as its associated time-scale is considerably longer than the duration of a rupture front passage. 

The interactions along the interface are modeled with the traction-at-split-node method \cite{andrews:1973}. Each node of one solid along the interface is paired with the node of the other solid at the same location. Tractions across the interface are coupled such that the nodes stick together and move as a pair. When interface tractions $\tau(x)$ reach local strength $\tau_p(x)$, the node pair is split. Each node can assume its own position, which leads to local slip $\delta$, \textit{i.e.,} tangential opening of the interface. As slip occurs, $\delta > 0$, local interface traction is equal to a strength $\tau (x) = \tau_s(x)$ that reduces with further slip. While frictional strength is generally a complex phenomenon that may depend on local slip, slip rate, and state, we apply a simple linear dependence on local slip
\begin{equation}
  \tau_s (\delta) = \left\{
    \begin{array}{ll}
      \tau_p - (\tau_p - \tau_r) \delta / d_c &\; \delta \leq d_c\\
      \tau_r &\; \delta > d_c
      \end{array}
      \right. ~.
\label{eq:cohesivelaw}
\end{equation}
This approach is equivalent to the cohesive-element method in crack propagation simulations but applied in the shear direction and with finite residual strength.
We use the following cohesive interface properties: interface peak strength $\tau_p - \tau_r = 1.62~\mathrm{MPa}$ and fracture energy $\Gamma = 2.5~\mathrm{J/m}^2$ that were inferred from our experiments with the dry interface. This results in a characteristic weakening length $d_c = 3.09~\mu\mathrm{m}$, as for a linear slip weakening law $ \Gamma=(\tau_p - \tau_r) d_c / 2 $.

We apply shear tractions along all in-plane exterior boundaries, including the vertical left and right boundary [see Fig.~\ref{fig:Simfig01}(f)]. This is different to the experimental setup but enables uniform shear stress distribution along the interface, which reduces complexity of our simulations without sacrificing fundamental equivalence with the experiments. Furthermore, it is important to note that this system is symmetric with respect to the interface. Therefore, the normal stress along the interface remains constant during entire simulations. This, obviously makes a potential dependence of local strength to contact pressure, as one would expect from Coulomb's friction law, irrelevant for this model. We utilize the system anti-symmetry to reduce computational cost. We apply anti-symmetry conditions at $y=0$, modeling only one block with an anti-symmetry traction-at-split-node method. Local slip thus is $\delta = 2u_x$. Furthermore, we apply symmetry boundary conditions at $z=0$ in three-dimensional simulations or use a plane-stress assumption (unless stated otherwise) in two-dimensional setups. Finally, a rupture front is nucleated through a seed crack that is created by locally reducing the interface strength over increasingly large area until the seed crack becomes unstable and propagates dynamically.

\subsection{Rupture propagation between infinitely thin plates}
\label{sec:2dsim}

\begin{figure}[t]
	\includegraphics [width=\textwidth] {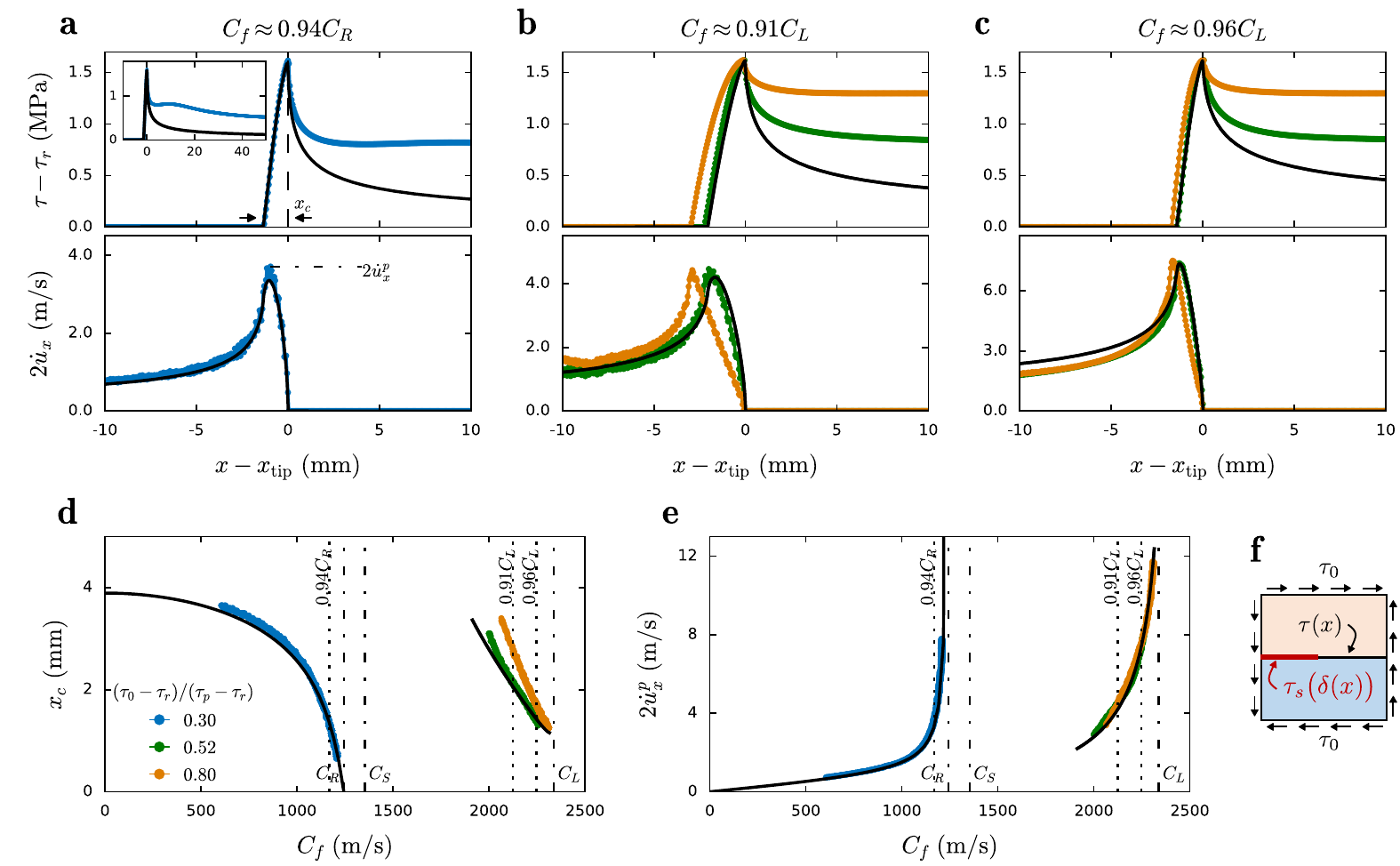}  
	\centering
	\caption{ Comparison of 2D plane stress simulations to LEFM predictions for  rupture speeds that range from sub-Rayleigh to supershear speeds. Three accelerating rupture events with different $\load$ values (denoted by different colors, see legend) are considered.	Shear traction $\sheartrac$ (top) and slip velocity $\slipvel$ (bottom) along the frictional interface ($y=0$) are presented at (a) sub-Rayleigh rupture speed $\Cf \approx 0.94 \CR$, (b) a supershear speed $\Cf \approx 0.91 \CL$ and (c) at $\Cf \approx 0.96 \CL$. Black lines - predictions of the steady state cohesive model with no adjustable parameters.  (inset in a) expanded range window of (a).	Systematic comparison of simulations to LEFM predictions of $\xc$ (d) and $\peakslipvel$ (e) for rupture speeds that range from slow to supershear speeds. $\xc$  contracts as $\Cf \rightarrow \CR$ and as $\Cf \rightarrow \CL$. $\peakslipvel$ diverges as $\Cf \rightarrow \CR$ and as $\Cf \rightarrow \CL$. (f) Schematic representation of numerical configuration with applied boundary conditions.}
	\label{fig:Simfig01}
\end{figure}

We start by comparing non-steady-state simulations with steady-state analytical cohesive zone model solutions. Our approach simulates infinitely thin plates (2D - plane stress), which was shown to be a good assumption for sub-Rayleigh fronts \cite{svetlizky2016}. Figure~\ref{fig:Simfig01}(a) presents typical sub-Rayleigh spatial profiles of the shear stress at the interface $\tau(x)$ and slip velocity $\slipvel$ ($\dot u_x$ denotes the fault parallel particle velocity component, at $y=0^+$).

The simulated fields are well-described by the steady-state LEFM solution (black lines), with no adjustable parameters; both the simulation and the analytical cohesive zone model employ same interfacial and material properties. For the cohesive zone model we adopt here $\widetilde{\tau}(\xi)=1 - (-\xi)^{a}$ with $a=1.4$. A different choice of the functional form $\widetilde{\tau}(\xi)$ would lead to similar results with only minor quantitative differences.

Note that ahead of the rupture tip, as in the measurements presented in Fig.~\ref{fig:SubRayleighFields}(c), some important differences in $\tau(x)$ appear between the theoretical and numerical calculations. These imply that the spatial range in which the LEFM solution is valid is larger for  the $\varepsilon_{xx}$ strain component (as represented by $\slipvel$)  [Fig.~\ref{fig:Simfig01}(a) - bottom panel] than for the shear stress component $\tau$ [Fig.~\ref{fig:Simfig01}(a) - top panel]. This should be expected as the analytical cohesive zone models describe the regularized zone and the singular K-dominance region [see Fig.~\ref{fig:Xc}(a)] and all far-field non-singular contributions, that naturally emerge in the simulations, are neglected. Non-singular contributions include the finite pre-stress level $\tau_0$ and the radiated stress-waves in the form of a stress peak that propagates ahead of the rupture tip [see inset in Fig.~\ref{fig:Simfig01}(a)] \cite{svetlizky2016}. These stress-waves can reach the strength of the frictional interface and facilitate a transition to the supershear regime. This mechanism, known as Burridge-Andrews mechanism, has been extensively discussed in the literature \cite{burridge1973,andrews1976,dunham2007,liu_Lapusta2008,bruhat2016,albertini:2017}.

The Burridge-Andrews mechanism predicts that the transition length rapidly increases with decreasing pre-stress level. To obtain supershear rupture fronts within the simulated spatial domain at relatively low pre-stress levels, we employ an alternative nucleation process which was discussed in detail in Ref.~\cite{kammer2018}. In short, we impose a seed crack that propagates slightly faster than $\CS$. Beyond a critical length, the imposed crack loses stability and spontaneously proceeds to propagate at $\Cf>\CS$.

Figure~\ref{fig:Simfig01}(b) compares two supershear rupture events at different pre-stress levels. We first note that increasing pre-stress levels associated with the supershear ruptures further decreases the region of validity ahead of the rupture tip of the LEFM solutions for $\tau$. This is demonstrated in Fig.~\ref{fig:Simfig01}(b,c) top panels and in the experimental results presented (Fig.~\ref{fig:SuperShearFields}), and should be expected following our previous discussion. Surprisingly, however, while the slip velocity profile for a rupture front propagating at moderate stress levels $\load=0.52$ (green curve) is well described by the cohesive zone model, deviations can be observed at higher pre-stress level (orange line). These include differences in the form of the $\slipvel(x)$ profile and increased cohesive zone size $\xc$ [see definition in Fig.~\ref{fig:Simfig01}(a)]. At higher propagation speeds (or, equivalently, longer crack lengths), see Fig.~\ref{fig:Simfig01}c, these discrepancies significantly decrease; both $\tau(x)$ and $\slipvel(x)$ spatial profiles gradually approach the steady-state prediction of the cohesive zone model. These observations are summarized in Figure~\ref{fig:Simfig01}(d) and \ref{fig:Simfig01}(e) by systematically tracking $\xc$ and the peak slip velocity $2\dot{u}_{x}^p$ [see definition in Fig.~\ref{fig:Simfig01}(a)]. 

Previous work \cite{huang:2002} has shown that the stress intensity factor of a suddenly arrested supershear crack gradually approaches its equilibrium value, in striking contrast to a sub-Rayleigh case in which the stress intensity factor is established instantaneously. This result prevents the construction of non-uniformly propagating supershear cracks by the superposition of uniform propagation solutions, as is typically done in the sub-Rayleigh regime \cite{freund1972II}. Similarly, our simulations show that while the fields within the K-dominance region (and cohesive zone) associated with sub-Rayleigh rupture fronts are established almost instantaneously, the fields within the vicinity of supershear rupture tips are formed gradually and may depend on external loading conditions and rupture propagation history. 

While the processes in which the near-tip fields of supershear ruptures are established are of fundamental importance, the magnitude of the effects discussed in this section is insufficient to describe the discrepancies observed between our experiments and LEFM solutions. 
  
\subsection{Effect of sample thickness}
\label{sec:3dsim}

\begin{figure}[h]
	\includegraphics [width=0.5\textwidth] {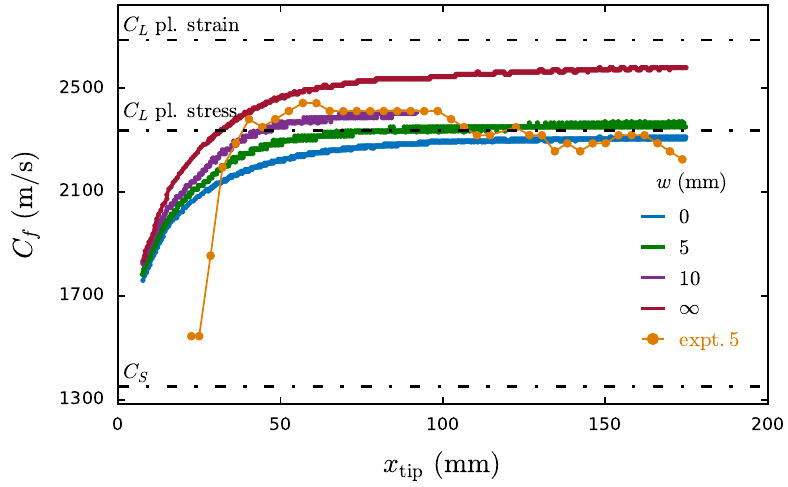}
	\centering
	\caption{Effects of the sample width $w$ on the limiting rupture speed. 2D and 3D supershear rupture simulations for $\load=0.8$ are considered. $w=0$ and $w=\infty$ correspond to 2D plane stress and plane strain simulations, respectively, while $w=5$ and $10$~mm cases are computed by complete 3D simulations. After nucleation, ruptures rapidly accelerate towards their terminal velocity $\Cf \rightarrow \CL^\mathrm{eff}$. $\CL^\mathrm{eff}$ depends on the relative width of the system $w/\xc$, ranging from plane stress to plane strain limiting cases (black dash-dotted lines). These 3D simulations provide an explanation for the experimental observations in which $\Cf>\CL^{\mathrm{plane \ stress}}$ (event 5 in Fig.~\ref{fig:Fig1}) that is shown here in orange for comparison.
	}
	\label{fig:Simfig02}
\end{figure}

The 2D plane-stress simulations, presented in the previous section, model the experimental system as infinitely thin. In this case, the longitudinal wave speed $\CL$ associated with the plane stress assumption provides a limiting rupture speed. In event 5 in Fig.~\ref{fig:Fig1}(c) and Fig.~\ref{fig:SuperShearFields}, however, we presented an example of a rupture that surpasses the predicted $\CL$ (plane-stress) value. Note that the rupture is still slower than the plain-strain value of $\CL$ (infinitely thick plates). This suggests that the finite thickness of the samples plays an important role in the supershear regime. In order to test this conjecture, we provide full 3D simulations with finite sample widths $w = 5~\mathrm{mm}$ and  $10~\mathrm{mm}$ (see Fig.~\ref{fig:Fig1} for the definition of $w$). Figure~\ref{fig:Simfig02} compares the rupture speed profiles obtained from our 3D and 2D simulations. The 2D simulations include plane-stress ($w = 0$) as well as plane-strain ($w = \infty$) assumptions and show that, after nucleation, ruptures accelerate and approach asymptotically their respective limiting speed $\CL$. In the 3D case, however, ruptures propagate at speeds that are bounded from below (plane stress) and above (plane strain) by the 2D simulations. It thus becomes obvious that ruptures propagating in a system with thin but finite plate thicknesses accelerate beyond $\CL^\mathrm{plane stress}$ and approach an effective longitudinal wave speed $\CL\mathrm{eff} > \CL^\mathrm{plane stress}$. This explains why event 5, shown again in orange in Fig.~\ref{fig:Simfig02}, temporarily exceeds $\CL^\mathrm{plane stress}$.

\begin{figure}
	\includegraphics[width=\textwidth] {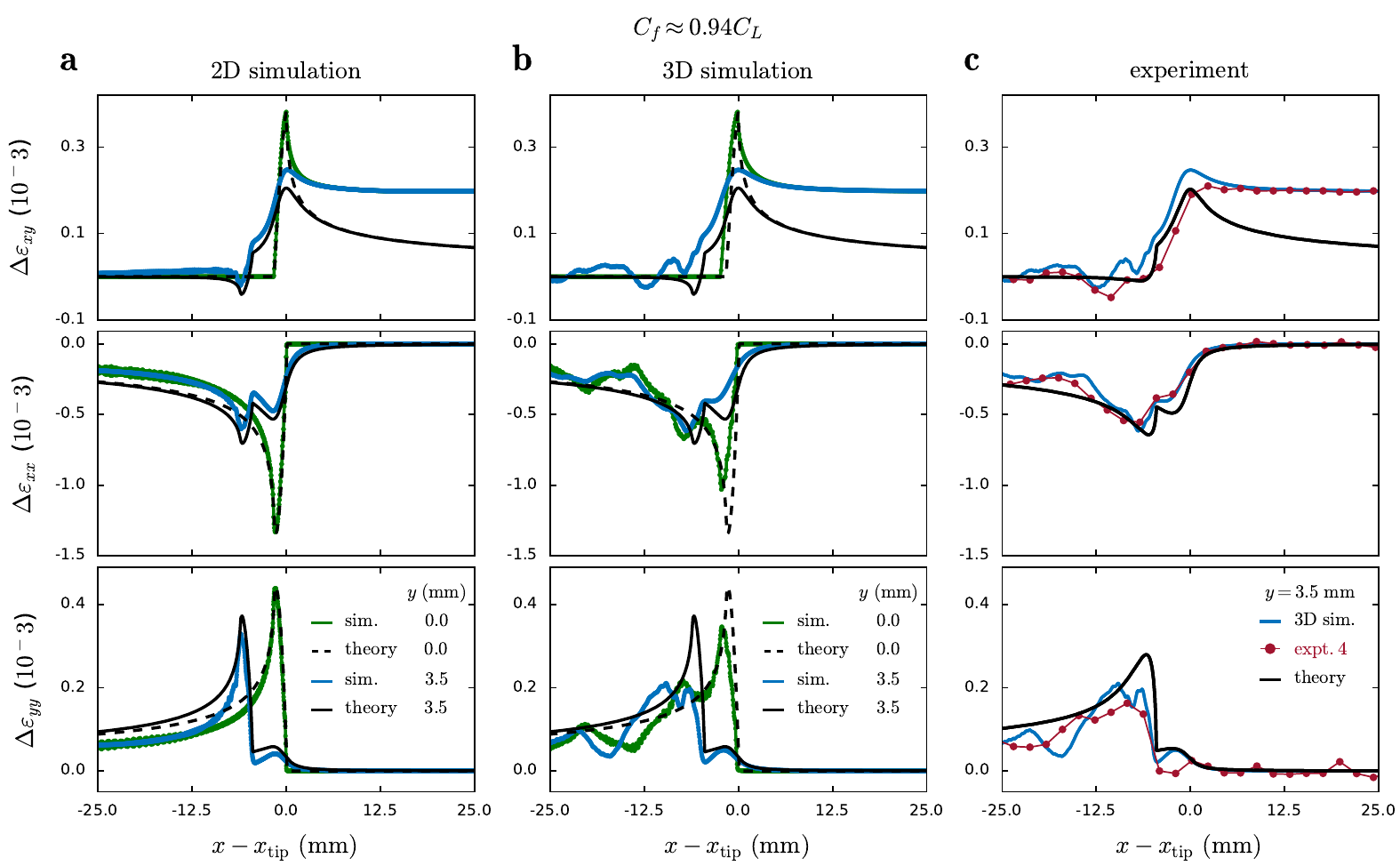}
	\centering
	\caption{Effects of the sample width on elastic strain fields at the supershear rupture tip vicinity. (a,b) $\strainij$ profiles at the interface ($y=0$ - green) and slightly above ($y=3.5$~mm - blue ) are plotted for moderate pre-stress level $\load= 0.52$. 2D plane stress simulation (a) and  3D simulation ($w=5$~mm) (b)  are compared to cohesive zone model predictions, plotted in dashed ($y=0$ ) and solid ($y=3.5$~mm) black lines. (c) Experimental measurements compared to 3D simulations and LEFM predictions. All simulations and LEFM predictions assume values for the cohesive zone properties, $\tau_p-\tau_r$ and $\Gamma$, as determined experimentally for the dry interface (Fig.~\ref{fig:SubRayleighFields} and Fig.~\ref{fig:Xc}). While LEFM predictions describe well on and off-fault fields observed in the 2D simulations (a), 3D simulations (b) reveal a complex structure of the fields due to reflections through the width of the sample. Experimental strain measurements during event 4 (c) are quantitatively reproduced by the 3D simulation, while the 2D simulation and the LEFM theory would overestimate the amplitudes of $\strainyy$ while agreeing well with $\strainxx$. For the 3D simulation $\strainij$ are reported at the sample face ($z=w/2$), corresponding to strain measurement locations in the experimental setup. Note that, to describe our simulations (a,b) and experiments (c), in the cohesive zone model we employ different functional forms of $\widetilde{\tau}(\xi)$, which results in a slightly different spatial functional form of strains (see text for more details).}
	\label{fig:Simfig03}
\end{figure}

How does the finite-thickness of the plates affect the near-tip strain fields? To address this question, Fig.~\ref{fig:Simfig03}(a) and Fig.~\ref{fig:Simfig03}(b) present strain fields during supershear rupture propagation obtained by the 2D and 3D simulations, respectively. We include here both the strain fields on ($y=0^+$) and away ($y=3.5$~mm) from the frictional interface for moderate pre-stress levels [$\load=0.52$]. This same example was considered in Fig.~\ref{fig:Simfig01}. Strain fields obtained in 2D simulations, both on and away from the frictional interface, are well described by the cohesive zone model. The effects of finite sample thickness on the structure of the near-tip strain are evident in Fig.~\ref{fig:Simfig03}(b). Although the main features remain generally similar, the analytical LEFM predictions and the 2D simulations fail to describe faithfully the detailed structure of the fields that emerges in cases of rupture fronts propagating in a 3D medium. Our 3D simulations reveal (1) significantly lower $\strainxx$ and $\strainyy$ amplitudes and, unexpectedly, (2) strong oscillations that are excited behind the rupture tip $x - \xtip < 0$. 

Furthermore, Fig.~\ref{fig:Simfig03}c demonstrates that the 3D simulations compare quantitatively well with the experiments. In particular, our 3D simulations describe well both the measured strain amplitudes and the apparent oscillations.
The origin of the strain oscillations are waves that follow the rupture tip and are reflected within $z$ direction from one free surface to the other. These waves do not occur in 3D simulations of sub-Rayleigh rupture (not shown here) as all elastic wave speeds are faster than the crack and thus every wave precedes the rupture. Free surfaces, however, can have other effects on sub-Rayleigh ruptures and promote the transition to the supershear regime \cite{kaneko:2010}. Detailed analysis of these trailing waves of supershear ruptures is left for future work.

\begin{figure}[h]
	\includegraphics[width=\textwidth]{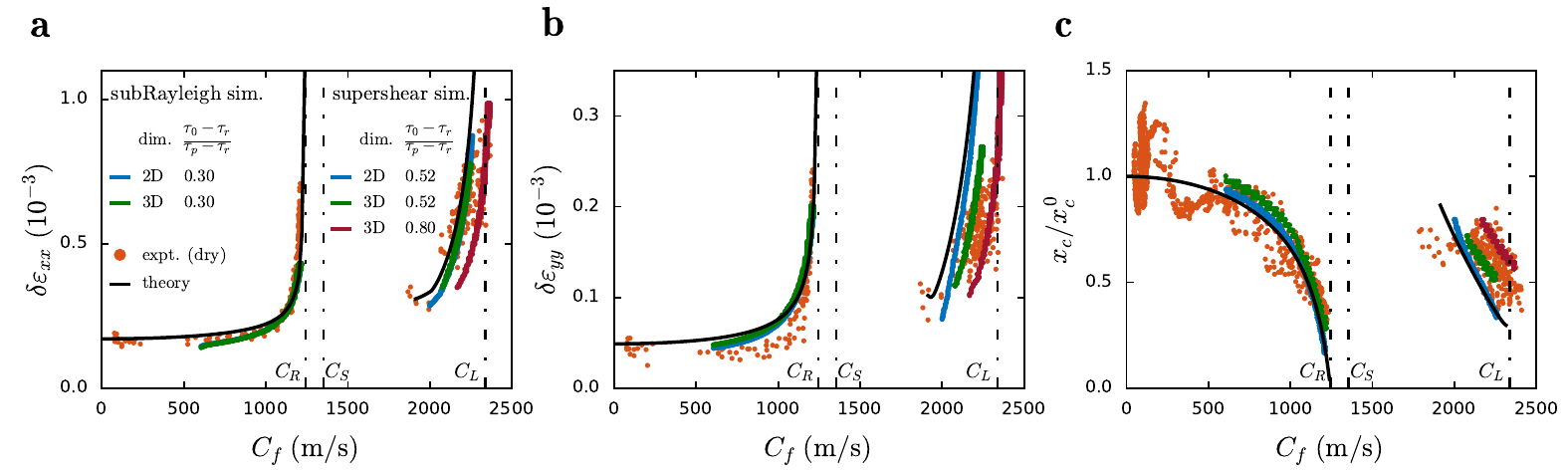}
	\centering
	\caption{
		Systematic comparison between experiments (dry case), 2D and 3D simulations and LEFM analytical predictions. Elastic strains, measured at $y=3.5$~mm, are characterized by their amplitudes $\strainampxx$ (a) and $\strainampyy$ (b) (see definitions in Fig.~\ref{fig:SubRayleighFields} and Fig.~\ref{fig:SuperShearFields}). (c) $\xc(\Cf)$ were normalized by $\xc^0=\xc(\Cf=0)$ to provide quantitative comparison as different definitions of $\xc$ were used for the experiments and simulations (see text for more details). (a-c) 	$\Gamma$ and $\tau_p-\tau_r$  were determined during slow rupture propagation and then used in the simulations and LEFM predictions across the whole range of $\Cf$. While for $\Cf<\CR$ 2D approximation of the experiments is justified, for $\Cf>\CS$ 3D description is necessary to capture the spread in the measurements. $\load$ values for the simulations are indicated in the caption. For the experiments $0.3<\load<0.8$ for $\Cf>\CS$. 
        }
	\label{fig:Simfig04}
\end{figure}

Figure~\ref{fig:Simfig04} provides a systematic comparison between the experiments (dry case), numerical simulations (both 2D and 3D) and the analytical LEFM predictions. Peak strain values $\strainampxx$ and $\strainampyy$ (see definition in Fig.~\ref{fig:SuperShearFields}) and the cohesive zone size $\xc$ are plotted with respect to the rupture speed. We normalize $\xc$ by $\xc^0=\xc(\Cf=0)$, to provide quantitative comparison between experiments and simulations. This is required as different functional forms of $\widetilde{\tau}(\xi)$ and definitions of $\xc$ were used [comparing Fig.~\ref{fig:Xc}(c) with Fig.~\ref{fig:Simfig01}(a)].

For $\Cf<\CR$ there is a good overall agreement between simulations, experiments and theory. This observation indicates that, in the sub-Rayleigh regime, the 2D approximation of the thin plates in our experiment is well justified. Within the supershear regime, however, this approximation shows some limitations and a 3D description is required for good quantitative agreement. The spread observed in our measurements at $\Cf>\CS$ is well captured by the 3D simulations. In particular, while the 3D effects are negligible for moderate loading (compare blue and green curves), for high $\load$, the finite-thickness clearly affects strain amplitudes and $\xc$ values.

\section{Discussion and Conclusion}
\label{sec:discussionconclusion}

Although natural earthquakes have long been conceptually studied within the fracture-mechanics framework \cite{book:scholz:2002,poliakov2002,bhat2007,xu2015,xu2016,xu2017}, direct experimental evidence that LEFM can provide quantitative predictions for frictional failure has started to accumulate only recently. In particular, experiments in thermoplastics \cite{svetlizky2014,mello2016} and rocks \cite{kammer:2019,xu:2019} have demonstrated that the universal square-root singular solutions, originally developed for brittle shear cracks, indeed describe the stress fields in vicinity of sub-Rayleigh rupture tips. 

Less is known, however, about how these singular fields are regularized \cite{rubino2017} and to what extent the fracture mechanics framework can be applied at the supershear propagation regime. Interestingly, these questions are inherently coupled. A striking contrast exists between sub-Rayleigh and supershear ruptures. Once a square-root singular region in the vicinity of a sub-Rayleigh rupture tip is established, LEFM can be used to provide quantitative predictions for rupture acceleration \cite{svetlizky2017a}, rupture arrest \cite{kammer2015,bayart2016a,ke:2018} and stress-wave radiation \cite{svetlizky2016}. These were previously reviewed in Ref.~\cite{svetlizky2019}. In the supershear propagation regime, however, near-tip stresses, stress-wave radiation \cite{broberg1999,mello2010,mello2016} and the rupture equation of motion \cite{broberg1994,kammer2018} depend explicitly on how the singular stresses are regularized at the rupture tip. Here, we provide estimates of how the singular fields are regularized and extend the applicability of the fracture-mechanics framework to the supershear rupture propagation regime by showing that analytical and numerical fracture-mechanics solutions provide quantitative description of the near-tip fields. 

Previous spatiotemporal analysis of supershear ruptures \cite{mello2010,mello2016} demonstrated a similarity between the measured fault-normal and fault-parallel particle velocities components and fracture-mechanics predictions. In particular, it was shown that the measured jump in particle velocities, associated with the Mach front, obeys speed dependent scaling deduced from steady-state dispersion relations. Information about any of the cohesive zone properties within these relations, however, scales out, and thus cannot be determined from the experimental measurements. By direct comparison of full 2D strain tensor and real contact area we show here that the cohesive zone model successfully captures the measurements across the whole rupture speed range and directly provides estimates for local interface properties. 

The first important property, the fracture energy $\Gamma$, which is initially determined through analysis of near-tip strain fields of slow ruptures, remains rupture-speed independent across the entire sub-Rayleigh and supershear speed range. The second property, the peak frictional strength $\tau_p$, is much more challenging to measure. In the sub-Rayleigh regime, the near-tip strain fields at the strain gauge locations are characterized by singular fields and do not contain any information about $\tau_p$. We thus infer $\tau_p$ from estimates of the cohesive zone size, which are provided by the real contact area measurements. At supershear speeds, however, shock waves reach locations far from the interface, and their amplitude depends on $\tau_p$. Strain measurements during the propagation of supershear rupture fronts thus contain direct information about the cohesive zone and provide an alternative approach to estimates of $\tau_p$. We find that both methods, based on contact area and strain fields, result in consistent values for the value of $\tau_p$. As a final validation of the inferred values of $\Gamma$ and $\tau_p$ we note that these exact values, for the same set of experiments, had been used in our prior work to describe successfully the equation of motion of sub-Rayleigh \cite{svetlizky2017a} and supershear \cite{kammer2018} rupture fronts.

While analytical fracture-mechanics solutions describe many aspects of frictional rupture fronts well across the entire speed range, some discrepancies appear for $\Cf>\CS$. In particular, the strain amplitudes $\strainampxx$ and $\strainampyy$ are over-estimated by the analytical solution. Our dynamic simulations capture well these discrepancies and illustrate that transient and geometrical (finite thickness of elastic medium) effects are the cause for these observations. Similar Mach front attenuation in 3D setups was observed in Ref.~\cite{dunham:2008a} and reduced ground motion amplitude was shown to occur near the free-surface of simulated supershear earthquakes \cite{zhang:2017}. Additionally, once the finite-thickness of the plates is taken into account, our simulations show that the limiting rupture speed is larger than the plane-stress approximation. This observation  explains why in high pre-stress experiments frictional rupture fronts exceeded slightly $\CL^\mathrm{plane stress}$. This is the first explicit demonstration of the effects of finite thickness of the interface, relative to the often-used assumption of a 2D system in describing rupture dynamics and mechanics. In this sense, these results are significant. First, they both validate the fact that the 2D assumption provides a good qualitative description of rupture dynamics across the entire range of rupture velocities. Second, these calculations denote important quantitative effects that, until now, have only been the subject of speculation. These effects include the limiting velocities of ruptures together with phenomena  such as the significantly increased noise that, we find, is inherent in the 3D signals.

Considering the onset of frictional sliding as a dynamic fracture process which is quantitatively described by fracture-mechanics  analytical solutions and numerical simulations is far from a trivial result. Some fundamental differences exist between ``standard'' tensile cracks and frictional rupture fronts. While cracks break material, rupture fronts only cause relative displacement between two already distinct bodies. More importantly, cracks faces are traction-free but rupture fronts continuously dissipate energy through frictional sliding at their tail, and can even propagate multiple times through the same interface. For these reasons, providing quantitative experimental  validation for the fracture-mechanics paradigm for frictional rupture fronts is crucial. To model our experiments we adopt in our numerical simulations the slip-weakening constitutive law of friction. This is the simplest class of cohesive zone model that we could have chosen; frictional resistance weakens to a residual value $\tau_r$ which is independent of rate. The use of this class of models is justified by previous observations in Ref.~\cite{svetlizky2017b}, where only slight rate dependence of $\tau_r$ was reported for PMMA. This observation enables us to map frictional rupture fronts to cracks by employing the linearity of the governing equations and subtracting $\tau_r$. Interestingly, however, recent work \cite{barras2019a,barras2019b} demonstrated the emergence of crack-like behavior of frictional rupture fronts for more realistic rate dependent friction. The simplicity of the adopted model (slip-weakening) enables us to highlight the importance of non-steady propagation (section~\ref{sec:2dsim}) and geometrical effects (section~\ref{sec:3dsim}) present during propagation of supershear rupture fronts. 

\section*{Acknowledgments}
We acknowledge the support of the Israel Science Foundation (Grants 1523/15 and 840/19). I.S. acknowledges the USIEF Fulbright program.

\appendix

\section{Rosette strain gage calibration}

We use miniature Vishay 015RJ rosette strain gages for local strain measurements. Each rosette strain gage is composed of three active regions. Each active region provides a strain component, $\varepsilon_i$, along the directions denoted by the white arrows in Fig.~\ref{fig:Appendix}(a).

Electrical resistance strain gages can be calibrated to a high precision when used on stiff materials such as various metals. However, when these strain gages are embedded on low modulus materials such as plastics, their presence might locally alter the strain field in their surroundings (see \cite{Ajovalasit2013} and references within). Analytical models and numerical data exist in the literature to enable us to estimate this effect and properly calibrate the measurement of strain. One convenient strain configuration that can be used to calibrate the strain gages is that of a disk under compression by a point load.

For purposes of calibration, a rosette strain gage was glued at the center of a $100$~mm diameter PMMA disk ($7.5$~mm width). The disk was subjected to radial compression at various angles, $\theta$, with respect to the rosette axis [$y'$ axis in Fig. \ref{fig:Appendix}(a)]. For this stress configuration the full stress field can be calculated analytically \cite{Timoshenko1990} and can, therefore, be used to perform a quantitative calibration of the strain gages. We assumed that a transformation that relates the altered strain field due to the rosette presence (here denoted by $\varepsilon_i'$) to the ``actual'' strain field in its absence ($\varepsilon_i$) could be found. We indeed found that the calibration measurements can be described by a phenomenological transformation of the following form:
\begin{eqnarray}
  \varepsilon_1'=a_1\cdot\varepsilon_1+k_1\cdot\varepsilon_3+g_1\cdot\varepsilon_{x''y''} 
  \label{eq:ShearSensTransformationBasic1}\\
  \varepsilon_2'=a_2\cdot\varepsilon_2+k_2\cdot\varepsilon_{x'x'}+g_2\cdot\varepsilon_{x'y'}\\
  \varepsilon_3'=a_3\cdot\varepsilon_3+k_3\cdot\varepsilon_1+g_3\cdot\varepsilon_{x''y''}
  \label{eq:ShearSensTransformationBasic3}
\end{eqnarray}
Where $a_i$ are corrections for the gage factors. $k_i$ and $g_i$ are the effective transverse and shear sensitivities of each of the gage components, respectively. (${x'',y''}$) is a coordinate system rotated by $45^{\circ}$ relative to the $(x',y')$ [see Fig. \ref{fig:Appendix}(a)]. 

We choose $a_1=a_3=1$ as only the relative calibration of the components is of interest. Due to reflection symmetry with respect to the $y'$ axis,  $k_1=k_3=k$, $g_1=g_3=g$ and $g_2=0$. This reflection symmetry does not exist with respect to  $\varepsilon_2$ and $\varepsilon_3$ - and hence effective shear sensitivity can not be excluded. To our knowledge, the effects of the elastic mismatch of the rosette configuration have not been previously considered. Furthermore, we note that shear sensitivity has not been discussed in the literature. Here, we find that shear sensitivity exists and is crucial for proper gage calibration. Taking these considerations into account and by using the relations
\begin{eqnarray}
  \varepsilon_{x''y''}=1/2(\varepsilon_1+\varepsilon_3)-\varepsilon_2\\
  \varepsilon_{x'x'}=\varepsilon_1+\varepsilon_3-\varepsilon_2
\end{eqnarray}
for $\varepsilon_{x''y''}$ and $\varepsilon_{x'x'}$ that appear in \ref{eq:ShearSensTransformationBasic1}-\ref{eq:ShearSensTransformationBasic3} we get
\begin{equation}
  \left(\begin{array}{ccc} 1+g/2 & -g & k+g/2 \\ k_2 & a_2-k_2 & k_2 \\ k+g/2 & -g & 1+g/2 \end{array}\right)
  \left(\begin{array}{c} \varepsilon_1 \\ \varepsilon_2 \\\varepsilon_3 \end{array}\right) = \left(\begin{array}{c} \varepsilon_1' \\ \varepsilon_2' \\\varepsilon_3' \end{array}\right)
  \label{eq:ShearSensTransformation}
\end{equation}
Once $\varepsilon_i'$ are measured the $\varepsilon_i$ can be calculated by using the inverse transformation. Our calibrations for the micromeasurements rosettes used in our experiments (Vishay 015RJ) mounted on PMMA revealed that $a_2\approx 0.95$, $k_2\approx-0.08$, $k\approx0$ and $g\approx0.1$.

\section*{Details of calibration}

A disk under ideal point loading has a complete analytical solution \cite{Timoshenko1990}. At the disk center, when the $y$ axis is aligned with the direction of compression, the strain tensor is given by
\begin{equation}
  \varepsilon_{ij}= \frac{P/E}{\pi t R} \left(\begin{array}{cc} 1+3\nu&0\\ 0&-(3+\nu)  \end{array} \right)
  \label{eq:DisckStrain}
\end{equation}
where $P$ is the applied force, $E(\approx3.2$~GPa) is the Young's modulus, $\nu(\approx0.33)$ is the Poisson's ratio and $t$ and $R$ denote the width and the radius of the disk, respectively. 

The four free parameters ($a_2,k_2,k,g$) that appear in the transformation \ref{eq:ShearSensTransformation} were determined by pressing the disk at various values of $\theta$. First, $k$ was determined by pressing the disk at $\theta=\pm45^{\circ}$. $a_2$ and $k_2$ were determined by pressing the disk at $\theta=0$ and $\theta=90^{\circ}$. In both cases the free parameters were deduced by comparing the measurements with the predictions given by \ref{eq:DisckStrain} ($\varepsilon_{xx}\approx-0.6\varepsilon_{yy}$). 

\begin{figure}
  \includegraphics[width=\textwidth]{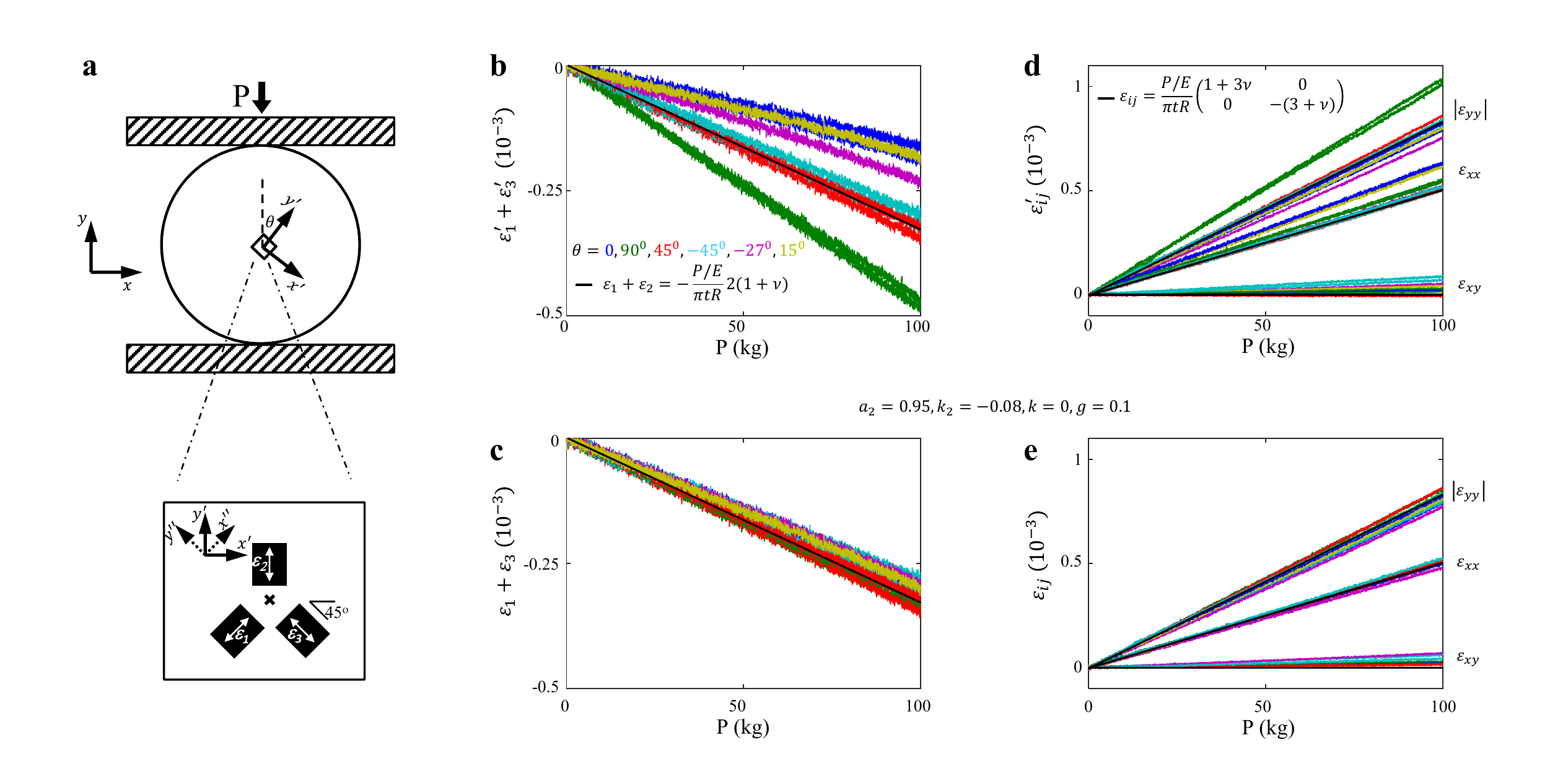}
  \centering
  \caption{\label{fig:Appendix}
    (a) Disk under compression at various angles, $\theta$, with respect to the rosette axis ($y'$ axis). The black rectangles represent the active area of the measuring components, while the white arrows represent the direction of the measured strains  $\varepsilon_1',\varepsilon_2',\varepsilon_3'$. (b-e) Colors correspond to different values of $\theta$. Black lines - theoretical prediction. Measurements of $\varepsilon_{1}'+\varepsilon_{3}'$ (b) and $\varepsilon_{ij}'$ (d) reveal a considerable spread for different values of $\theta$. Corrected values (using Eq.~ \ref{eq:ShearSensTransformation}) of $\varepsilon_{1}+\varepsilon_{3}$ (c) and $\varepsilon_{ij}$ (e)  collapse to the theoretical predictions. The small discrepancies that appear in $\varepsilon_{xy}$  (e), we believe, are due to a $\sim2^{\circ}$ uncertainty in $\theta$. The values of $\varepsilon_{j}'$, $\varepsilon_{ij}'$ and $P$ at $P=50$~kg were subtracted to account for initial misalignment of the disk.
  }
\end{figure}

We are now in a position to determine the shear sensitivity $g$. Note that the trace of the strain tensor is invariant under rotations of the coordinate system. Figure \ref{fig:Appendix}(b) demonstrates, however, that the measured values of $\varepsilon_1'+\varepsilon_3'$ depend on $\theta$ and therefore do not measure the trace, as one would naively assume. Importantly, no value of the transverse sensitivity can account for these discrepancies. Measurements for all values of $\theta$ collapse well to a single line, predicted by Eq.~\ref{eq:DisckStrain}, when the shear sensitivity is taken into account ($g\approx0.1$) and $\varepsilon_1+\varepsilon_3$ are calculated according to Eq.~\ref{eq:ShearSensTransformation} (see Fig.~\ref{fig:Appendix}(c)). 

Finally, we summarize the results in figures \ref{fig:Appendix}(d) and \ref{fig:Appendix}(e). In Fig.~\ref{fig:Appendix}(d) we use the measured $\varepsilon_{j}'$ to calculate $\varepsilon_{ij}'$ (in the $(x,y)$ coordinate system) with no corrections assumed. Note the apparent spread. Figure~\ref{fig:Appendix}(e) demonstrates that once the transformation given by \ref{eq:ShearSensTransformation} was used to calculate $\varepsilon_{ij}$, the spread disappears and the results agree well with the theoretical predictions.

\section*{References}

\bibliography{cited}

\end{document}